\def\DOFINAL{1} 

\ifx\DOFINAL\undefined
	\documentclass[preprint,amsmath,linenumbers,endfloats]{revtex4-1} 
	\gdef\mybreak{} \gdef\mybreakt{} \gdef\myqquad{} 
\else
	\documentclass[reprint,amsmath]{revtex4-1} 
	\gdef\mybreak{\notag\\} \gdef\mybreakt{\times\notag\\} \gdef\myqquad{\qquad}  
\fi

\usepackage[utf8]{inputenc}
\usepackage{siunitx} 
\usepackage[]{natbib}
\usepackage{mathtools} 
\usepackage{booktabs} 
\usepackage{xspace} 
\DeclareMathAlphabet{\mathpzc}{OT1}{pzc}{m}{it}
\usepackage{calc}    
\usepackage{xstring} 
\usepackage{intcalc} 
\usepackage{tikz}    
\usepackage{tkz-euclide} 
\usepackage{graphicx}
\usepackage[]{subcaption} 

\newcommand{\drawtriad}[7]{ 
	
	\xdef\compA{#2}
	\xdef\compB{#3}
	\xdef\compC{#1}
	\xdef\offsetx{#4}
	\xdef\offsety{#5}
	\xdef\offsetpoint{#6}	
	\xdef\angRot{#7}

	\pgfmathsetmacro{\angA}{acos(( (\compC)^2.0 +(\compB)^2.0 - (\compA)^2.0)/(2.0*\compC*\compB))}
	\pgfmathsetmacro{\angC}{acos(( (\compA)^2.0 +(\compB)^2.0 - (\compC)^2.0)/(2.0*\compA*\compB))}
	\pgfmathsetmacro{\angB}{180.0-\angA-\angC}

	\ifnum\offsetpoint=1 
		\path (\offsetx,\offsety) coordinate (A) ++(\angA+\angRot:\compC) coordinate (B) ++(-\angC+\angRot:\compA) coordinate (C);
	\else \relax \fi	

	\ifnum\offsetpoint=2
		\path (\offsetx,\offsety) coordinate (B) ++(-\angC+\angRot:\compA) coordinate (C) ++(180.0+\angRot:\compB) coordinate (A);
	\else \relax \fi		
	
	\ifnum\offsetpoint=3
		\path (\offsetx,\offsety) coordinate (C) ++(180.0+\angRot:\compB) coordinate (A) ++(\angA+\angRot:\compC) coordinate (B);	
	\else \relax \fi		

	\draw[\arwstyleA] (A)--(B) node[midway, fill=white] {\arwlabelA};
	\draw[\arwstyleB] (B)--(C) node[midway, fill=white] {\arwlabelB};
	\draw[\arwstyleC] (C)--(A) node[midway, fill=white] {\arwlabelC};

	\ifnum\rotAround=1 
		\pgfmathsetmacro{\angBpr}{180.0-\angB}
		\pgfmathsetmacro{\radius}{\compA*sin(\angBpr)}
		\pgfmathsetmacro{\len}{\compA*cos(\angBpr)}
		\coordinate (E) at ($(B)+(\angA+\angRot:\len)$);
		\pgfmathsetmacro{\rott}{\angA-90}
	    \pgfmathsetmacro{\radiusy}{\radius/7} 
	    \draw[dashed, thick, rotate around={\angRot+\rott:(A)}] (E) ellipse (\radius cm and \radiusy cm);	
	\else \relax \fi	

	\ifnum\rotAround=2
		\pgfmathsetmacro{\angApr}{-90+\angB}
     	\pgfmathsetmacro{\radius}{\compB*cos(\angA+\angApr)}
		\coordinate (E) at ($(A)+(\angA+\angApr+\angRot:\radius)$);
		\pgfmathsetmacro{\rott}{\angA+\angApr}
	    \pgfmathsetmacro{\radiusy}{\radius/7} 
	    \draw[dashed, thick, rotate around={\angRot+\rott:(E)}] (E) ellipse (\radius cm and \radiusy cm);	
	\else \relax \fi		
	
	\ifnum\rotAround=3
	    \pgfmathsetmacro{\len}{\compC*cos(\angA)} 
		\pgfmathsetmacro{\radius}{\compC*sin(\angA)} 
		\coordinate (E) at ($(A)+(\angRot:\len)$);
		\pgfmathsetmacro{\rott}{90.0}
	    \pgfmathsetmacro{\radiusy}{\radius/7} 
	    \draw[dashed, thick, rotate around={\angRot+\rott:(E)}] (E) ellipse (\radius cm and \radiusy cm);	
	\else \relax \fi		
}

\newcommand{\intactclass}[1]
{	
	\xdef\theclass{#1} 	
 	
	\begin{tikzpicture}[
      triad/.style={-latex new, arrow head=2.2mm, thick},
      dominant/.style={line width=3pt,->},
      subdom/.style={line width=1.8pt,->},
      scale=1.4
    ]

	\xdef\angRot{-90}
	\xdef\compA{1.5^2}
	\xdef\compB{1.5^3}
	\xdef\compC{1.5^1}
	
	\pgfmathsetmacro{\angA}{acos((+(\compC)^2 +(\compB)^2 - (\compA)^2)/(2*\compC*\compB))}
	\pgfmathsetmacro{\angC}{acos((-(\compC)^2 +(\compB)^2 + (\compA)^2)/(2*\compA*\compB))}
	\pgfmathsetmacro{\angB}{180-\angA-\angC}

	\pgfmathsetmacro{\triangleAx}{0} 
	\pgfmathsetmacro{\triangleAy}{0} 
	\pgfmathsetmacro{\triangleBx}{\compC*cos(\angA+\angRot)} 
	\pgfmathsetmacro{\triangleBy}{\compC*sin(\angA+\angRot)} 
	\pgfmathsetmacro{\triangleCx}{\compB*cos(\angRot)} 
	\pgfmathsetmacro{\triangleCy}{\compB*sin(\angRot)} 

	\coordinate (A) at (\triangleAx, \triangleAy);
	\coordinate (B) at (\triangleBx, \triangleBy);
	\coordinate (C) at (\triangleCx, \triangleCy);

	\coordinate (Am) at ($(A)!0.5!(B)$);
	\coordinate (Bm) at ($(B)!0.5!(C)$);
	\coordinate (Cm) at ($(C)!0.5!(A)$);

	\ifnum\theclass=1 
		\draw[triad] (A)--(B) node[midway,right] {$u_{n}^+$};
		\draw[triad] (B)--(C) node[midway,right] {$u_{n+p}^-$};
		\draw[triad] (C)--(A) node[midway,left]  {$u_{n+q}^+$};
		\path [dominant,bend right=10]  (Am) edge (Bm);
		\path [subdom,bend left=10]   (Am) edge (Cm);
	\else \relax \fi	

	\ifnum\theclass=2
		\draw[triad] (A)--(B) node[midway,right] {$u_{n}^+$};
		\draw[triad] (B)--(C) node[midway,right] {$u_{n+p}^-$};
		\draw[triad] (C)--(A) node[midway,left]  {$u_{n+q}^-$};
		\path [subdom,bend left=10]    (Bm) edge (Am);
		\path [dominant,bend right=10] (Bm) edge (Cm);
	\else \relax \fi	
	
	\ifnum\theclass=3 
		\draw[triad] (A)--(B) node[midway,right] {$u_{n}^+$};
		\draw[triad] (B)--(C) node[midway,right] {$u_{n+p}^+$};
		\draw[triad] (C)--(A) node[midway,left]  {$u_{n+q}^-$};
		\path [dominant,bend right=10]  (Am) edge (Bm);
		\path [dominant,bend left=10] (Am) edge (Cm);
	\else \relax \fi	

	\ifnum\theclass=4 
		\draw[triad] (A)--(B) node[midway,right] {$u_{n}^+$};
		\draw[triad] (B)--(C) node[midway,right] {$u_{n+p}^+$};
		\draw[triad] (C)--(A) node[midway,left]  {$u_{n+q}^+$};
		\path [dominant,bend left=10]  (Bm) edge (Am);
		\path [subdom,bend right=10] (Bm) edge (Cm);
	\else \relax \fi	

	\end{tikzpicture}
}

\newenvironment{helicalmodelalternative}[6]
{
    
    \begin{tikzpicture}[
      helmodepos/.style={},
      helmodeneg/.style={dashed},
      shellline/.style={}, 
      pointsigma/.style={draw,circle,fill=white,inner sep=0pt,minimum size=20pt},
      pointlocked/.style={draw,circle,fill=white,inner sep=0pt,minimum size=20pt},
      dominant/.style={line width=2.8pt},
      subdom/.style={line width=1.3pt},
      rev/.style={color={rgb:red,228;green,26;blue,28},->-},
      fwd/.style={color={rgb:red,55;green,126;blue,184},->-},
      scale=0.7
    ]

\tikzset{->-/.style={decoration={
  markings,
  mark=at position .7 with {\arrow[very thick]{triangle 90}}},postaction={decorate}}}

\ifx\DOFINAL\undefined
    \def \shpad {5em}
    \def \shheight {16.7em}
    \def \shstart {\shheight}
\else
    \def \shpad {6.5em}
    \def \shheight {16.7em}
    \def \shstart {\shheight}
\fi

    \xdef\mySp{#1}
    \xdef\mySq{#2}
    
	\draw[shellline] ({1*\shpad},\shstart) -- ({1*\shpad},\shheight) node[above] {{${n-2}$}};
	\draw[shellline] ({2*\shpad},\shstart) -- ({2*\shpad},\shheight) node[above] {{${n-1}$}};
	\draw[shellline] ({3*\shpad},\shstart) -- ({3*\shpad},\shheight) node[above] {{${n}$}};
	\draw[shellline] ({4*\shpad},\shstart) -- ({4*\shpad},\shheight) node[above] {{${n+1}$}};
	\draw[shellline] ({5*\shpad},\shstart) -- ({5*\shpad},\shheight) node[above] {{${n+2}$}};
	
    \def \offset {1em}	
    \def \pair {2em}
    \def \interpair {5.5em}
	
	\foreach \s in {1,2,3}
	{
		\draw[helmodepos] ({0.7*\shpad},{\offset + (\s-1)*\interpair + \pair}) -- 
		                  ({5.4*\shpad},{\offset + (\s-1)*\interpair + \pair}) node[right] {$u^{+}$};
		\draw[helmodeneg] ({0.7*\shpad},{\offset + (\s-1)*\interpair + 0em})    -- 
						  ({5.4*\shpad},{\offset + (\s-1)*\interpair + 0em})    node[right] {$u^{-}$};
	}

	\gdef\Sk{}

	\coordinate (A) at ({1*\shpad},{\offset + #5*\pair + (2)*\interpair});
	\coordinate (B) at ({2*\shpad},{\offset + #6*\pair + (2)*\interpair});
	\coordinate (C) at ({3*\shpad},{\offset +    \pair + (2)*\interpair});

	\ifnum\mySp=1
		\ifnum\mySq=1
		\draw [dominant,rev]  (B) to [bend right=70] (A);
    	\draw [subdom,fwd]   (B) to [bend left=70] (C); 	
   	    \else \relax \fi
    \else \relax \fi

	\ifnum\mySp=0
		\ifnum\mySq=0
			\draw [subdom,rev]    (B) to [bend right=70] (A);
	    	\draw [dominant,fwd]   (B) to [bend left=70] (C); 	
   	    \else \relax \fi
    \else \relax \fi
    
	\ifnum\mySp=1
		\ifnum\mySq=0
			\draw [dominant,fwd] (A) to [bend left=40] (B);
	    	\draw [dominant,fwd] (A) to [bend left=40] (C); 	
   	    \else \relax \fi
    \else \relax \fi
    
	\ifnum\mySp=0
		\ifnum\mySq=1
			\draw [dominant,fwd]  (A) to [bend left=40] (B);
	    	\draw [subdom,fwd]   (A) to [bend left=20] (C); 	
   	    \else \relax \fi
    \else \relax \fi

	\draw node[pointsigma]  at (A) {\footnotesize{$\sigmaE$}};
	\draw node[pointsigma]  at (B) {\footnotesize{$\sigmaF$}};
    \draw node[pointlocked] at (C) {\footnotesize{$+$}};

	\coordinate (A) at ({2*\shpad},{\offset + #3*\pair + (1)*\interpair});
	\coordinate (B) at ({3*\shpad},{\offset +    \pair + (1)*\interpair});
	\coordinate (C) at ({4*\shpad},{\offset + #4*\pair + (1)*\interpair});
	
	\ifnum\mySp=1
		\ifnum\mySq=1
		\draw [dominant,rev]  (B) to [bend right=70] (A);
    	\draw [subdom,fwd]   (B) to [bend left=70] (C); 	
   	    \else \relax \fi
    \else \relax \fi

	\ifnum\mySp=0
		\ifnum\mySq=0
			\draw [subdom,rev]  (B) to [bend right=70] (A);
	    	\draw [dominant,fwd]   (B) to [bend left=70] (C); 	
   	    \else \relax \fi
    \else \relax \fi
    
	\ifnum\mySp=1
		\ifnum\mySq=0
			\draw [dominant,fwd]  (A) to [bend left=-40] (B);
	    	\draw [dominant,fwd]   (A) to [bend left=-40] (C); 	
   	    \else \relax \fi
    \else \relax \fi
    
	\ifnum\mySp=0
		\ifnum\mySq=1
			\draw [dominant,fwd]  (A) to [bend left=-40] (B);
	    	\draw [subdom,fwd]   (A) to [bend left=-20] (C); 	
   	    \else \relax \fi
    \else \relax \fi

	\draw node[pointsigma]  at (A) {\footnotesize{$\sigmaC$}};
    \draw node[pointlocked] at (B) {\footnotesize{$+$}};
	\draw node[pointsigma]  at (C) {\footnotesize{$\sigmaD$}};

	\coordinate (A) at ({3*\shpad},{\offset +    \pair + (0)*\interpair});
	\coordinate (B) at ({4*\shpad},{\offset + #1*\pair + (0)*\interpair});
	\coordinate (C) at ({5*\shpad},{\offset + #2*\pair + (0)*\interpair});

	\ifnum\mySp=1
		\ifnum\mySq=1
		\draw [dominant,rev]  (B) to [bend right=70] (A);
    	\draw [subdom,fwd]   (B) to [bend left=70] (C); 	
   	    \else \relax \fi
    \else \relax \fi

	\ifnum\mySp=0
		\ifnum\mySq=0
			\draw [subdom,rev]  (B) to [bend right=70] (A);
	    	\draw [dominant,fwd]   (B) to [bend left=70] (C); 	
   	    \else \relax \fi
    \else \relax \fi
    
	\ifnum\mySp=1
		\ifnum\mySq=0
			\draw [dominant,fwd]  (A) to [bend left=-40] (B);
	    	\draw [dominant,fwd]   (A) to [bend left=-40] (C); 	
   	    \else \relax \fi
    \else \relax \fi
    
	\ifnum\mySp=0
		\ifnum\mySq=1
			\draw [dominant,fwd]  (A) to [bend left=30] (B);
	    	\draw [subdom,fwd]   (A) to [bend left=20] (C); 	
   	    \else \relax \fi
    \else \relax \fi
    
    \draw node[pointlocked] at (A) {\footnotesize{$+$}};
	\draw node[pointsigma]  at (B) {\footnotesize{$\sigmaA$}};
	\draw node[pointsigma]  at (C) {\footnotesize{$\sigmaB$}};
    
	\gdef\Sk{s}
}
{
    \end{tikzpicture}\\
}

\AtBeginDocument{
\heavyrulewidth=.08em
\lightrulewidth=.05em
\cmidrulewidth=.03em
\belowrulesep=.65ex
\belowbottomsep=0pt
\aboverulesep=.4ex
\abovetopsep=0pt
\cmidrulesep=\doublerulesep
\cmidrulekern=.5em
\defaultaddspace=.5em
}

\ifx\DOFINAL\undefined

\fi

\def\suppmaterial{appendix\xspace}

\def\submodel{submodel\xspace}
\def\Submodel{Submodel\xspace}
\def\submodels{submodels\xspace}

\def\submodelweight{submodel weight\xspace}
\def\submodelweights{submodel weights\xspace}
\def\coupledmodel{coupled model\xspace}

\def\pqset{\lbrace p,q \rbrace\xspace}
\def\SpSqset{\lbrace \Sp,\Sq \rbrace\xspace}

\renewcommand{\Re}{\operatorname{Re}}

\gdef\NLderiv{d_{t}\big\vert_{\mathrm{N.L.}}}

\usepackage{ifthen} 

\newcommand{\eqnref}[1]{(\ref{#1})} 
\newcommand{\vect}[1]{\mathbf{#1}} 
\newcommand{\I}{\imath} 

\gdef\genexpEsymb{\alpha}
\gdef\genexpHsymb{\beta}
\gdef\pseudosymb{\dagger}

\gdef\genexpE{\genexpEsymb_i}

\gdef\genexpH{\genexpHsymb_i}

\newcommand{\genE}[1]{\lambda^{#1} } 
\newcommand{\genH}[1]{\lambda^{#1} } 

\gdef\Einvar#1{E^{(#1)}}
\gdef\Einvarreal{E}
\gdef\Einvarpseudo{E^{\pseudosymb}}
\gdef\Hinvar#1{H^{(#1)}}
\gdef\Hinvarreal{H}
\gdef\Hinvarpseudo{H^{\pseudosymb}}

\gdef\Eflux#1#2{\Pi^{\Einvar{#1}}_{#2}}
\gdef\Efluxreal#1{\Pi^{\Einvarreal}_{#1}}

\gdef\Hflux#1#2{\Pi^{\Hinvar{#1}}_{#2}}
\gdef\Hfluxreal#1{\Pi^{\Hinvarreal}_{#1}}

\gdef\tempE{A}
\gdef\temppseudoE{A^{\pseudosymb}}
\gdef\tempH{B}
\gdef\temppseudoH{B^{\pseudosymb}}

\newcommand{\MYVAR}[5]{ 
	\ifthenelse{\equal{#2}{}}{\vect{#1}}{#1_{#2}}^{#3}
	\ifthenelse{\equal{#4}{}}{}{(\ifthenelse{\equal{#5}{}}{\vect{#4}}{#4_{#5}})} 
}
\newcommand{\hs}[3]    {\vect{h}_{#1}^{#2}\ifthenelse{\equal{#3}{}}{}{(\vect{#3})}} 
\newcommand{\hslong}[3]{\vect{h}_{#1}^{#2}(#3)} 
\newcommand{\triadgeom}[9]{ 
	      \vect{h}^{#2}_{#1}(#3)
	\times\vect{h}^{#5}_{#4}(#6)
	\cdot \vect{h}^{#8}_{#7}(#9) 
} 

\newcommand{\uu}[4]{\MYVAR{u}{#1}{#2}{#3}{#4}} 
\newcommand{\uf}[2]{\uu{#1}{}{#2}{}}  
\newcommand{\sprime}{^{\prime}\mkern-1.2mu}
\newcommand{\dprime}{^{\prime\prime}\mkern-1.2mu}

\newcommand{\uhel}[3]{u_{#1}^{#2}(#3)}
\gdef\USH#1#2{u_{#1}^{#2}}

\newcommand{\ush}[3]{ \ifthenelse{\equal{#3}{1}}{u_{#1}^{#2,*} }{ u_{#1}^{#2} } }

\gdef\Sk{s}
\gdef\Sp{s\sprime}
\gdef\Sq{s\dprime}

\gdef\VK{\vect{k}}
\gdef\VP{\vect{k\sprime}}
\gdef\VQ{\vect{k\dprime}}

\gdef\VKLEN{k}
\gdef\VPLEN{k\sprime}
\gdef\VQLEN{k\dprime}

\gdef\VKHAT{\vect{\hat{k}}}
\gdef\VPHAT{\vect{\hat{k}\sprime}}
\gdef\VQHAT{\vect{\hat{k}\dprime}}

\newcommand{\vk}[1]{
	\ifthenelse{\equal{#1}{1}}{ \vect{k}}{
	\ifthenelse{\equal{#1}{2}}{ \vect{k}}{	
	\ifthenelse{\equal{#1}{3}}{ \vect{k}}{
	\ifthenelse{\equal{#1}{-1}}{k}{
	\ifthenelse{\equal{#1}{-2}}{k}{
	\ifthenelse{\equal{#1}{-3}}{k}{
	}}}}}}
} 
\newcommand{\vp}[1]{
	\ifthenelse{\equal{#1}{0}}{ \vect{k\sprime}}{
	\ifthenelse{\equal{#1}{1}}{\vp{0}}{
	\ifthenelse{\equal{#1}{2}}{\vp{0}}{	
	\ifthenelse{\equal{#1}{3}}{\vp{0}}{
	\ifthenelse{\equal{#1}{-1}}{k_{}\sprime}{
	\ifthenelse{\equal{#1}{-2}}{k_{}\sprime}{
	\ifthenelse{\equal{#1}{-3}}{k_{}\sprime}{
	}}}}}}}
} 
\newcommand{\vq}[1]{
	\ifthenelse{\equal{#1}{0}}{ \vect{k\dprime}}{
	\ifthenelse{\equal{#1}{1}}{\vq{0}}{
	\ifthenelse{\equal{#1}{2}}{\vq{0}}{	
	\ifthenelse{\equal{#1}{3}}{\vq{0}}{
	\ifthenelse{\equal{#1}{-1}}{k_{}\dprime}{
	\ifthenelse{\equal{#1}{-2}}{k_{}\dprime}{
	\ifthenelse{\equal{#1}{-3}}{k_{}\dprime}{
	}}}}}}}
} 

\newcommand{\vkhatrot}[1]{
	\ifthenelse{\equal{#1}{1}}{\VKHAT}{
	\ifthenelse{\equal{#1}{2}}{\VPHAT}{
	\ifthenelse{\equal{#1}{3}}{\VQHAT}{
	}}}
} 
\newcommand{\vphatrot}[1]{
	\ifthenelse{\equal{#1}{1}}{\VPHAT}{
	\ifthenelse{\equal{#1}{2}}{\VKHAT}{
	\ifthenelse{\equal{#1}{3}}{\VKHAT}{
	}}}
} 
\newcommand{\vqhatrot}[1]{
	\ifthenelse{\equal{#1}{1}}{\VQHAT}{
	\ifthenelse{\equal{#1}{2}}{\VQHAT}{
	\ifthenelse{\equal{#1}{3}}{\VPHAT}{
	}}}
} 

\newcommand{\vkrotrel}[1]{
	\ifthenelse{\equal{#1}{1}}{\VK}{
	\ifthenelse{\equal{#1}{2}}{\lambda^{-p}\VP}{
	\ifthenelse{\equal{#1}{3}}{\lambda^{-q}\VQ}{
	\ifthenelse{\equal{#1}{-1}}{\VKLEN}{
	\ifthenelse{\equal{#1}{-2}}{\VPLEN\lambda^{-p}}{
	\ifthenelse{\equal{#1}{-3}}{\VQLEN\lambda^{-q}}{
	}}}}}}
} 
\newcommand{\vprotrel}[1]{
	\ifthenelse{\equal{#1}{1}}{\VP}{
	\ifthenelse{\equal{#1}{2}}{\lambda^{-p}\VK}{
	\ifthenelse{\equal{#1}{3}}{\lambda^{-q}\VK}{
	\ifthenelse{\equal{#1}{-1}}{\VPLEN}{
	\ifthenelse{\equal{#1}{-2}}{\VKLEN\lambda^{-p}}{
	\ifthenelse{\equal{#1}{-3}}{\VKLEN\lambda^{-q}}{
	}}}}}}
} 
\newcommand{\vqrotrel}[1]{
	\ifthenelse{\equal{#1}{1}}{\VQ}{
	\ifthenelse{\equal{#1}{2}}{\lambda^{-p}\VQ}{
	\ifthenelse{\equal{#1}{3}}{\lambda^{-q}\VP}{
	\ifthenelse{\equal{#1}{-1}}{\VQLEN}{
	\ifthenelse{\equal{#1}{-2}}{\VQLEN\lambda^{-p}}{
	\ifthenelse{\equal{#1}{-3}}{\VPLEN\lambda^{-q}}{
	}}}}}}
} 

\newcommand{\vkrotabs}[1]{
	\ifthenelse{\equal{#1}{1}}{\vkrotabs{-1}\vkhatrot{1}}{
	\ifthenelse{\equal{#1}{2}}{\vkrotabs{-2}\vkhatrot{2}}{
	\ifthenelse{\equal{#1}{3}}{\vkrotabs{-3}\vkhatrot{3}}{
	\ifthenelse{\equal{#1}{-1}}{\VKLEN}{
	\ifthenelse{\equal{#1}{-2}}{\VKLEN}{
	\ifthenelse{\equal{#1}{-3}}{\VKLEN}{
	}}}}}}
} 
\newcommand{\vprotabs}[1]{
	\ifthenelse{\equal{#1}{1}}{\vprotabs{-1}\vphatrot{1}}{
	\ifthenelse{\equal{#1}{2}}{\vprotabs{-2}\vphatrot{2}}{
	\ifthenelse{\equal{#1}{3}}{\vprotabs{-3}\vphatrot{3}}{
	\ifthenelse{\equal{#1}{-1}}{\lambda^p\VKLEN}{
	\ifthenelse{\equal{#1}{-2}}{\lambda^{-p}\VKLEN}{
	\ifthenelse{\equal{#1}{-3}}{\lambda^{-q}\VKLEN}{
	}}}}}}
} 
\newcommand{\vqrotabs}[1]{
	\ifthenelse{\equal{#1}{1}}{\vqrotabs{-1}\vqhatrot{1}}{
	\ifthenelse{\equal{#1}{2}}{\vqrotabs{-2}\vqhatrot{2}}{
	\ifthenelse{\equal{#1}{3}}{\vqrotabs{-3}\vqhatrot{3}}{
	\ifthenelse{\equal{#1}{-1}}{\lambda^q\VKLEN}{
	\ifthenelse{\equal{#1}{-2}}{\lambda^{q-p}\VKLEN}{
	\ifthenelse{\equal{#1}{-3}}{\lambda^{p-q}\VKLEN}{
	}}}}}}
}

\newcommand{\krotidxshifted}[1]{
	\ifthenelse{\equal{#1}{1}}{\KIDX-q}{
	\ifthenelse{\equal{#1}{2}}{\KIDX-q+p}{
	\ifthenelse{\equal{#1}{3}}{\krotidx{3}}{
	}}}
}
\newcommand{\protidxshifted}[1]{
	\ifthenelse{\equal{#1}{1}}{\KIDX-q+p}{
	\ifthenelse{\equal{#1}{2}}{\KIDX-q}{
	\ifthenelse{\equal{#1}{3}}{\protidx{3}}{
	}}}
}
\newcommand{\qrotidxshifted}[1]{
	\ifthenelse{\equal{#1}{1}}{\KIDX}{
	\ifthenelse{\equal{#1}{2}}{\KIDX}{
	\ifthenelse{\equal{#1}{3}}{\qrotidx{3}}{
	}}}
}

\gdef\KIDX{n}
\newcommand{\krotidx}[1]{
	\ifthenelse{\equal{#1}{1}}{\KIDX}{
	\ifthenelse{\equal{#1}{2}}{\KIDX}{
	\ifthenelse{\equal{#1}{3}}{\KIDX}{
	\ifthenelse{\equal{#1}{-1}}{\KIDX-2}{
	\ifthenelse{\equal{#1}{-2}}{\KIDX-1}{
	\ifthenelse{\equal{#1}{-3}}{\KIDX}{
	}}}}}}
}
\newcommand{\protidx}[1]{
	\ifthenelse{\equal{#1}{1}}{\KIDX+p}{
	\ifthenelse{\equal{#1}{2}}{\KIDX-p}{
	\ifthenelse{\equal{#1}{3}}{\KIDX-q}{
	\ifthenelse{\equal{#1}{-1}}{\KIDX-1}{
	\ifthenelse{\equal{#1}{-2}}{\KIDX-2}{
	\ifthenelse{\equal{#1}{-3}}{\KIDX-2}{
	}}}}}}
}
\newcommand{\qrotidx}[1]{
	\ifthenelse{\equal{#1}{1}}{\KIDX+q}{
	\ifthenelse{\equal{#1}{2}}{\KIDX+q-p}{
	\ifthenelse{\equal{#1}{3}}{\KIDX+p-q}{
	\ifthenelse{\equal{#1}{-1}}{\KIDX}{
	\ifthenelse{\equal{#1}{-2}}{\KIDX}{
	\ifthenelse{\equal{#1}{-3}}{\KIDX-1}{
	}}}}}}
}

\gdef\eqsgn#1#2{   \ifthenelse{\equal{#1}{#2}}{+}{-}}
\gdef\eqsgninv#1#2{\ifthenelse{\equal{#1}{#2}}{-}{+}}

\newcommand{\effsign}[2]{
	\StrLeft{#1}{1}[\firstletterA]
	\StrLeft{#2}{1}[\firstletterB]
	\ifthenelse{\equal{\firstletterA}{\firstletterB}}{+}{-}
}
\newcommand{\effsym}[2]{
	\effsign{#1}{#2}\StrGobbleLeft{#2}{1}
}
\newcommand{\effsymnp}[2]{
	\StrLeft{#1}{1}[\firstletterA]
	\StrLeft{#2}{1}[\firstletterB]
	\StrLen{#2}[\lengthB]
	\ifthenelse{\equal{\firstletterA}{\firstletterB}}{
		\ifthenelse{\equal{\lengthB}{1}}{+}{} 
	}{-} 
	\StrGobbleLeft{#2}{1}
}

\gdef\WGfunc#1{G_{#1}}
\gdef\WG{\WGfunc{p,q}}
\gdef\WGdef{1/8\,(2\lambda^{-2q} + 2\lambda^{-2p} + 2 -\lambda^{-2(p+q)} \mybreak &\myqquad\myqquad - \lambda^{2(p-q)} - \lambda^{2(q-p)})^{1/2}}

\gdef\Wgfunc#1#2{g_{#1}^{#2}}
\gdef\Wgfuncnorm#1#2{\hat{g}_{#1}^{#2}}
\gdef\Wg{\Wgfunc{p,q}{\Sp,\Sq}}
\gdef\Wglocal{\Wgfunc{}{\Sp,\Sq}}
\gdef\Wgdef{-\Sp\Sq(1 + \Sp\lambda^{p} - \Sq\lambda^{q})(\Sp\lambda^p - \Sq\lambda^q)}
\gdef\Wgdeflocal{-\Sp\Sq(1 + \Sp\lambda - \Sq\lambda^{2})(\Sp\lambda - \Sq\lambda^2)}

\gdef\Wepsfunc#1#2{\epsilon_{#1}^{#2}}
\gdef\Weps{\Wepsfunc{p,q}{\Sp,\Sq}}
\gdef\Wepslocal{\Wepsfunc{}{\Sp,\Sq}}
\gdef\Wepsdeffunc#1#2#3#4{\frac{1 - #4\lambda^{#2}}{\lambda^{#1}-#3#4\lambda^{#2}}}

\gdef\Wepsdef{\Wepsdeffunc{p}{q}{\Sp}{\Sq}}
\gdef\Wepsdeflocal{\Wepsdeffunc{}{2}{\Sp}{\Sq}}

\gdef\Wxifunc#1#2{\xi_{#1}^{#2}}
\gdef\Wxi{\Wxifunc{p,q}{\Sp,\Sq}}
\gdef\Wxilocal{\Wxifunc{}{\Sp,\Sq}}
\gdef\Wxideffunc#1#2#3{-#3(1- #2\epsilon_{#1}^{#2,#3})}
\gdef\Wxidef{\Wxideffunc{p,q}{\Sp}{\Sq}}
\gdef\Wxideflocal{\Wxideffunc{}{\Sp}{\Sq}}

\gdef\Egeneqn#1#2#3{1 - \Sp\ifthenelse{\equal{#1}{1}}{\genE{#3}}{(\genE{#3})^{#1}}
\ifthenelse{\equal{#1}{1}}{\Wepsfunc{}{\Sp,\Sq}}{\Wepsfunc{#1,#2}{\Sp,\Sq}}
+ \Sq(\genE{#3})^{#2}
\ifthenelse{\equal{#1}{1}}{\Wxifunc{}{\Sp,\Sq}}{\Wxifunc{#1,#2}{\Sp,\Sq}}
}

\gdef\Hgeneqn#1#2#3{1 -    \ifthenelse{\equal{#1}{1}}{\genH{#3}}{(\genH{#3})^{#1}}
\ifthenelse{\equal{#1}{1}}{\Wepsfunc{}{\Sp,\Sq}}{\Wepsfunc{#1,#2}{\Sp,\Sq}}
+    (\genH{#3})^{#2}
\ifthenelse{\equal{#1}{1}}{\Wxifunc{}{\Sp,\Sq}}{\Wxifunc{#1,#2}{\Sp,\Sq}}
}

\gdef\helampslong#1#2#3#4#5#6{u_{#1}^{#2}(#3)u_{#4}^{#5}(#6)} 
\newcommand{\helampsshort}[6]{u_{#3}^{#1#2}u_{#6}^{#4#5}} 
\newcommand{\hamps}[5]{
	\ifthenelse{\equal{#1}{0}}{ \helampslong{#2}{ }{#3}{#4}{ }{#5} }{ 
	\ifthenelse{\equal{#1}{1}}{ \helampslong{#2}{*}{#3}{#4}{ }{#5} }{ 
	\ifthenelse{\equal{#1}{2}}{ \helampslong{#2}{*}{#3}{#4}{ }{#5} }{ 
	\ifthenelse{\equal{#1}{3}}{ \helampslong{#2}{ }{#3}{#4}{ }{#5} }{ 
	\ifthenelse{\equal{#1}{-1}}{ \helampsshort{#2}{,*}{#3}{#4}{}{#5} }{ 
	\ifthenelse{\equal{#1}{-2}}{ \helampsshort{#2}{,*}{#3}{#4}{}{#5} }{ 
	\ifthenelse{\equal{#1}{-3}}{ \helampsshort{#2}{}{#3}{#4}{}{#5} }{ 
	}}}}}}}
} 

\gdef\pqcond{0<p<q}
\gdef\pqveccond{k_n+k_{n+p} \ge k_{n+q}}
\gdef\pqveccondinv{k_n+k_{n+p} \le k_{n+q}}

\gdef\OuterSum{\sum_{\mathclap{\;\pqcond}}\WG\sum_{\mathclap{\Sp,\,\Sq}}\Wg}

\gdef\sigmaA{\Sk\Sp}
\gdef\sigmaB{\Sk\Sq}
\gdef\sigmaC{\Sk\Sp}
\gdef\sigmaD{\Sk\Sp\Sq}
\gdef\sigmaE{\Sk\Sq}
\gdef\sigmaF{\Sk\Sp\Sq}

\gdef\NSESMsingle{
\hamps{-1}{\sigmaA}{\protidx{1}}{\sigmaB}{\qrotidx{1}}  
\mybreak & 
- \frac{\Weps}{\lambda^p}    
\hamps{-2}{\sigmaC}{\protidx{2}}{\sigmaD}{\qrotidx{2}} 
+ \frac{\Wxi}{\lambda^q}  
\hamps{-3}{\sigmaE}{\protidx{3}}{\sigmaF}{\qrotidx{3}}}

\gdef\Dissp{D_n}
\gdef\DisspdefSS{\nu \VKLEN_n^2}
\gdef\DisspdefLS{\nu_L \VKLEN_n^{-8}}
\gdef\Disspdef{\DisspdefSS + \DisspdefLS}

\newcommand{\NSESM}{(d_t & + \Dissp)\ush{\KIDX}{\Sk}{0} = \Sk\VKLEN_n \sum_{\mathclap{0<p<q}}\WG \sum_{\Sp,\Sq}\Wg \Big(\NSESMsingle\Big)}

\gdef\NSESMsingleLocal{
\hamps{-1}{\sigmaA}{\KIDX+1}{\sigmaB}{\KIDX+2}  \mybreak
- \frac{\Wepslocal}{\lambda}    
\hamps{-2}{\sigmaC}{\KIDX-1}{\sigmaD}{\KIDX+1} 
+ \frac{\Wxilocal}{\lambda^2}  
\hamps{-3}{\sigmaE}{\KIDX-2}{\sigmaF}{\KIDX-1}}

\newcommand{\NSESMLocal}{(d_t + \Dissp)\ush{\KIDX}{\Sk}{0} = \Sk\VKLEN_n \sum_{\Sp,\Sq}\Wglocal \Big(\NSESMsingleLocal\Big)}

\gdef\KIDXs{n}
\newcommand{\tripleintA}[6]{
\ush{\ifthenelse{\equal{#1}{}}{\KIDXs}{\KIDXs#1}}{#4}{1}    
\ush{\ifthenelse{\equal{#1}{}}{\KIDXs+1}{\KIDXs#2}}{#5}{1} 
\ush{\ifthenelse{\equal{#1}{}}{\KIDXs+2}{\KIDXs#3}}{#6}{0}}
\newcommand{\tripleintB}[6]{
\ush{\ifthenelse{\equal{#1}{}}{\KIDXs-1}{\KIDXs#1}}{#4}{1}  
\ush{\ifthenelse{\equal{#1}{}}{\KIDXs}{\KIDXs#2}}{#5}{1}   
\ush{\ifthenelse{\equal{#1}{}}{\KIDXs+1}{\KIDXs#3}}{#6}{0}}
\newcommand{\tripleintC}[6]{
\ush{\ifthenelse{\equal{#1}{}}{\KIDXs-2}{\KIDXs#1}}{#4}{0}  
\ush{\ifthenelse{\equal{#1}{}}{\KIDXs-1}{\KIDXs#2}}{#5}{0} 
\ush{\ifthenelse{\equal{#1}{}}{\KIDXs}{\KIDXs#3}}{#6}{1}}

\newcommand{\tripleintAconj}[6]{
\ush{\ifthenelse{\equal{#1}{}}{\KIDXs}{\KIDXs#1}}{#4}{0}    
\ush{\ifthenelse{\equal{#1}{}}{\KIDXs+1}{\KIDXs#2}}{#5}{0} 
\ush{\ifthenelse{\equal{#1}{}}{\KIDXs+2}{\KIDXs#3}}{#6}{1}}
\newcommand{\tripleintCb}[6]{
\ush{\ifthenelse{\equal{#1}{}}{\KIDXs-2}{\KIDXs#1}}{#4}{1} 
\ush{\ifthenelse{\equal{#1}{}}{\KIDXs-1}{\KIDXs#2}}{#5}{1} 
\ush{\ifthenelse{\equal{#1}{}}{\KIDXs}{\KIDXs#3}}{#6}{0}}

\newcommand{\modalcontr}[1]{(\vert\ush{\KIDXs}{+}{0}\vert^2 #1 \vert\ush{\KIDXs}{-}{0}\vert^2)}
\newcommand{\triplehelampsshort}[9]{
u_{#1}^{#3\ifthenelse{\equal{#2}{}}{}{,#2}}
u_{#4}^{#6\ifthenelse{\equal{#5}{}}{}{,#5}}
u_{#7}^{#9\ifthenelse{\equal{#8}{}}{}{,#8}}
}

\gdef\USHvert#1#2{\vert \USH{#1}{#2}\vert^2}

\gdef\meanhelicity{\delta_{\mathrm{in}}}
\gdef\meanhelicitypseudo{\delta^\dagger_{\mathrm{in}}}
\gdef\meanenergy{\epsilon_{\mathrm{in}}}
\gdef\meanenergypseudo{\epsilon^\dagger_{\mathrm{in}}}
\gdef\meanErspec#1{\langle \Einvarreal_{#1} \rangle}
\gdef\meanEpspec#1{\langle \Einvarpseudo_{#1} \rangle}
\gdef\meanHrspec#1{\langle \Hinvarreal_{#1} \rangle}
\gdef\meanHpspec#1{\langle \Hinvarpseudo_{#1} \rangle} 

\begin{document}

\title{The role of helicity in triad interactions in 3D turbulence investigated in a new shell model}

\author{Nicholas M. Rathmann}
\email{rathmann@nbi.ku.dk}
\affiliation{Niels Bohr Institute, University of Copenhagen, Denmark}
\author{Peter D. Ditlevsen}
\email{pditlev@nbi.ku.dk}
\affiliation{Niels Bohr Institute, University of Copenhagen, Denmark}

\begin{abstract}
Fully developed homogeneous isotropic turbulence in 2D is fundamentally different from 3D.
In 2D, the simultaneous conservation of both energy and enstrophy in the inertial ranges of scales leads to a forward cascade of enstrophy and a reverse cascade of energy. 
In 3D, helicity, the integral of the scalar product of velocity and vorticity, is also an inviscid flow invariant along with kinetic energy. 
Unlike enstrophy, however, helicity does not block the cascade of energy to small scales.
Energy and helicity are not only globally conserved but also conserved within each non-linear triadic interaction between three plane waves in the spectral form of the Navier--Stokes equation (NSE). 
By decomposing each plane wave into two helical modes of opposite helicities each triadic interaction is split into a set of eight triadic interactions
between helical modes \citep{bib:waleffe1992nature}.
\citet{bib:biferale2012inverse} recently found that a subset of these interactions which render both signs of helicity separately conserved (i.e. enstrophy-like) leads to an inverse cascade of (part of) the energy. 
Motivated by this finding we introduce a new shell model obtained from the NSE expressed in the helical basis \citep{bib:waleffe1992nature}. By analysing and integrating the new model we attempt to explain why the dual forward cascade of energy and helicity dominates in 3D turbulence.
\end{abstract}

\date{\today}
\maketitle

\section{Introduction}
The role played by helicity in the cascade processes of fully developed 3D turbulence is elusive. 
Helicity, the integral of the scalar product of vorticity and velocity, is an inviscid invariant thought to be more or less passively advected through the energy 
cascade from the integral to the viscous scale of the flow. 
This stands in contrast to the case of 2D turbulence where enstrophy, the integral of vorticity squared, is a second positive inviscid invariant besides energy. 
The ratio of the dissipation 
of enstrophy to energy scales with the Kolmogorov scale $\eta$ as $\eta^{-2}$, thus for $\eta\rightarrow 0$ the 
forward cascade of enstrophy prevents a forward cascade of energy, which instead is transported to larger scales. 
Following \citet{bib:waleffe1992nature} we refer to this as a reverse cascade, synonymous to an inverse- or up-scale cascade.

A similar scaling argument for 3D turbulence leads the ratio of dissipation of helicity to energy
scaling as $\eta^{-1}$. 
Thus for a constant dissipation of helicity the dissipation of energy vanishes when $\eta\rightarrow 0$.
Unlike the 2D case, however, this does not prevent a forward cascade of energy because helicity is not sign specific. This implies that the
separate dissipation of positive and
negative helicity structures can grow as $\eta^{-1}$ while the net dissipation of both energy and helicity balance the input at the forcing scale.
In recent work by \citet{bib:biferale2012inverse} it was proposed that if only interactions among same-signed helicity modes are considered a phenomenon corresponding to the reverse energy cascade in 2D turbulence could be present in the 3D case with sign-fixed helicity playing the role of enstrophy. 

\begin{figure*}[]
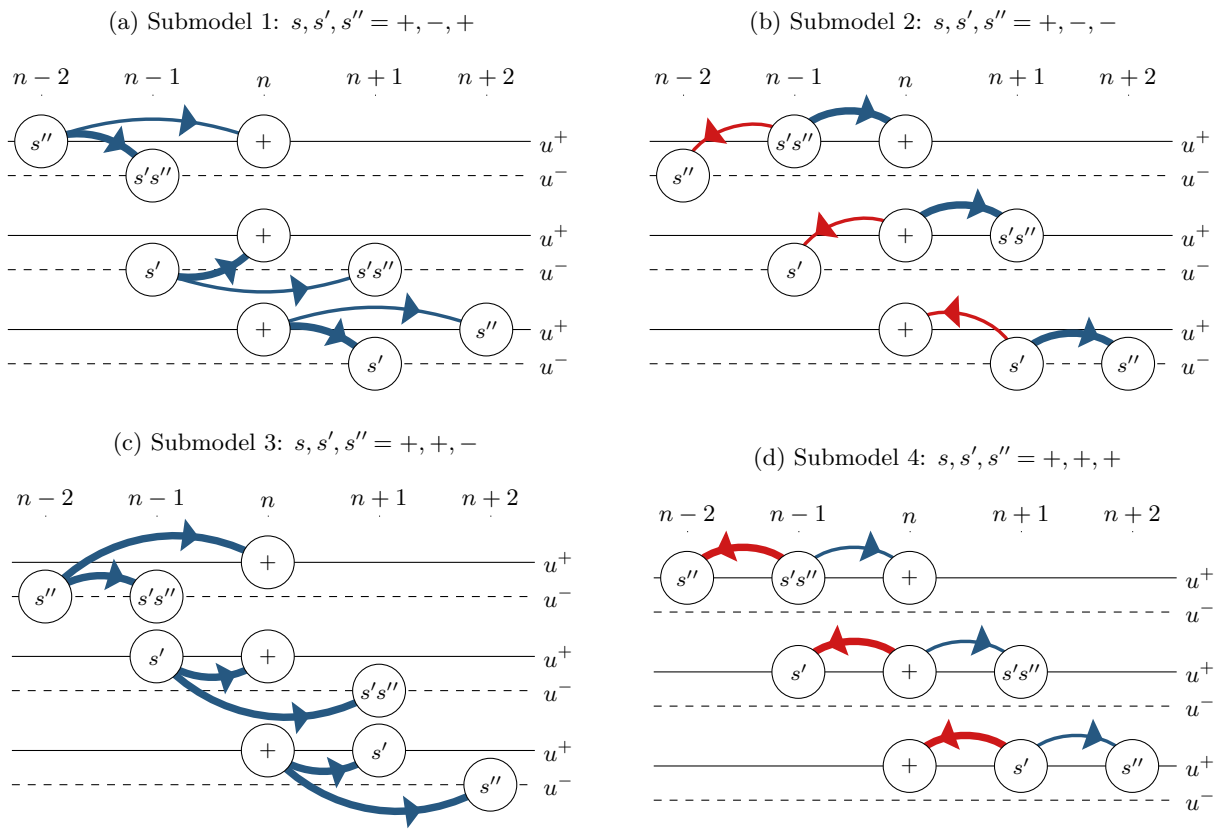

\centering
\begin{subfigure}{.47\textwidth}\caption{\Submodel 1: $\Sk,\Sp,\Sq = +,-,+$}\smallskip\centering\begin{helicalmodelalternative}{0}{1}{0}{0}{1}{0}\end{helicalmodelalternative}\end{subfigure}
\begin{subfigure}{.47\textwidth}\caption{\Submodel 2: $\Sk,\Sp,\Sq = +,-,-$}\smallskip\centering\begin{helicalmodelalternative}{0}{0}{0}{1}{0}{1}\end{helicalmodelalternative}\end{subfigure}
\bigskip\\
\begin{subfigure}{.47\textwidth}\caption{\Submodel 3: $\Sk,\Sp,\Sq = +,+,-$}\smallskip\centering\begin{helicalmodelalternative}{1}{0}{1}{0}{0}{0}\end{helicalmodelalternative}\end{subfigure}
\begin{subfigure}{.47\textwidth}\caption{\Submodel 4: $\Sk,\Sp,\Sq = +,+,+$}\smallskip\centering\begin{helicalmodelalternative}{1}{1}{1}{1}{1}{1}\end{helicalmodelalternative}\end{subfigure}
\caption[Helical couplings]{Schematic of the three non-linear helical interactions of each \submodel coupling to $\ush{n}{+}{0}$. All interactions are sign flipped for complementary interactions coupling to $\ush{n}{-}{0}$ (not shown). The arrows indicate the average energy transfer direction within each triadic interaction resulting from a linear stability analysis \citep{bib:waleffe1992nature}: blue arrows denote forward energy transfers while red denote reverse. The thick arrows represent the dominant transfers while the thin represent subordinate transfers.}\label{fig:helicalcouplings}
\end{figure*}

In the spectral representation of the Navier-Stokes equation (NSE) non-linear interactions are represented by exchanges of energy and helicity between three plane waves under the constraint that their wave vectors (momenta) sum to zero, thus forming triangles (triads). 
Within any such single triadic interaction between three waves conservation of energy and helicity holds.
In the interest of investigating the cascade of helicity in an incompressible flow it is useful to further decompose the spectral velocity components $\uf{}{k}$ in terms of helical modes.
Under the helical decomposition spectral velocity components $\uf{}{\VKLEN}{}$ are decomposed onto a plane perpendicular to $\VK$ using incompressibility $\VK\cdot\vect{u}(\VK) = 0$ such that $\vect{u}(\VK) = u_{+}(\VK)\vect{h}_+(\VK) + u_{-}(\VK)\vect{h}_-(\VK)$. The basis vectors $\vect{h}_\pm(\VK)$ are eigenvectors of the curl operator, i.e. $\I \VK \times \vect{h}_\pm(\VK) = \pm\VKLEN\vect{h}_\pm(\VK)$, leading to energy and helicity being given by 
\begin{alignat}{5}
E&=\sum_\VK  (\vert u_{+}(\VK) \vert ^2 + \vert u_{-}(\VK) \vert^2) 
\\
H&=\sum_\VK k(\vert u_{+}(\VK) \vert ^2 - \vert u_{-}(\VK) \vert^2), \label{eqn:NSE_E_and_H}
\end{alignat}
and the spectral form of the NSE being given by \citep{bib:waleffe1992nature}
\gdef\HNSEweight#1{(\Sp\VPLEN - \Sq \VQLEN) #1\;\hs{\Sp}{*}{\VP}\times\hs{\Sq}{*}{\VQ}\cdot\hs{\Sk}{*}{\VK}}
\begin{align}
(\partial_t + \nu k^2)\uf{\Sk}{k} = - 1/4 \sum\limits_{\mathclap{\VK+\VP+\VQ=0}\,\,\qquad} \sum_{\Sp,\Sq} \HNSEweight{\mybreakt} \;  \uu{\Sp}{*}{\VPLEN}{}\uu{\Sq}{*}{\VQLEN}{}\label{eqn:Helical_NSE} ,
\end{align}
where $\Sk,\Sp,\Sq = \pm1$ are helical signs. The inner sum indicates that each triadic interaction is now split into a set of $2^3=8$ distinct sub-interactions among the helical modes, which may be divided into the four sub-groups:  
$\lbrace \Sk, \Sp,\Sq\rbrace = \pm \lbrace +,-,+\rbrace,\pm \lbrace +,-,-\rbrace,\pm\lbrace +,+,-\rbrace,\pm\lbrace +,+,+\rbrace$. 
The interaction coefficient 
\begin{align}
\HNSEweight{}
\end{align}
will, for a given triad of waves $\VK+\VP+\VQ=0$, give the relative weights of the four different helical sub-interactions. 
By considering only the three terms in (\ref{eqn:Helical_NSE}) corresponding to a single interaction between modes $\uu{\Sk}{}{\VKLEN}{}, \uu{\Sp}{}{\VPLEN}{}$ and $\uu{\Sq}{}{\VQLEN}{}$, the linear stability of the fixed points $(\uu{\Sk}{}{\VKLEN}{},\uu{\Sp}{}{\VPLEN}{},\uu{\Sq}{}{\VQLEN}{})=(U_0,0,0), (0,U_0, 0), (0,0,U_0)$ may easily be calculated.
\citet{bib:waleffe1992nature} proposed that the energy transfer within triads might be determined by this fixed-point stability such that energy flows out of the unstable mode into the other two. 
By this rationale the above four sub-interactions may be split into two classes; one in which energy flows from the smallest wave mode (large scales) to the two larger wave modes (smaller scales), termed the "forward" class, and two for which the energy flows out of the middle mode and into the largest and smallest mode, termed the "reverse" class. 
Note here that sub-interactions between same-signed helical modes corresponding to the 2D turbulence case are
of the "reverse" class. Note also that the largest wave mode (smallest scale) is never an unstable mode. 

In fully developed turbulence it is not clear to what extent linear stability analysis is relevant, and more importantly, to what extent mixing between the 
four different sub-interactions is essential for the overall behaviour of the flow. 
Even if the flow by some strong symmetry constraints could be prepared in a maximally helical state (of only one helical sign), linear instability would make energy flow into modes of opposite sign, obeying the helicity conservation by creating an equal amounts of helicity of both signs in the process. In this work we thus seek to investigate numerically the relative importance 
of the four sub-interactions in 3D turbulence. Motivated by this we introduce a new helical shell model
inspired by \eqnref{eqn:Helical_NSE} allowing the different helical sub-interactions to be mixed. 
Helically decomposed shell models derived from the regular GOY \citep{bib:gledzer1973system} and Sabra \citep{bib:sabra_lvov,bib:sabra_ditlevsen} shell models have already previously been studied \citep{bib:biferale1995role,bib:benzi1996helical,bib:biferale1998helicityA,bib:biferale1998helicityB,bib:DePietro2015arXiv150806390D}. Applying the decomposition to shell models, four possible helical shell models may be constructed, each one corresponding to one of the four sub-interactions among helical modes.  
Our new shell model, however, is structurally closer to \eqnref{eqn:Helical_NSE} and contains the coupling strengths for the four types of sub-interactions which are naturally derived from the NSE. In the following we will therefore refer to these as the four \submodels of the new helical shell model.

\begin{figure*}[!t]
\centering
\includegraphics[scale=0.88]{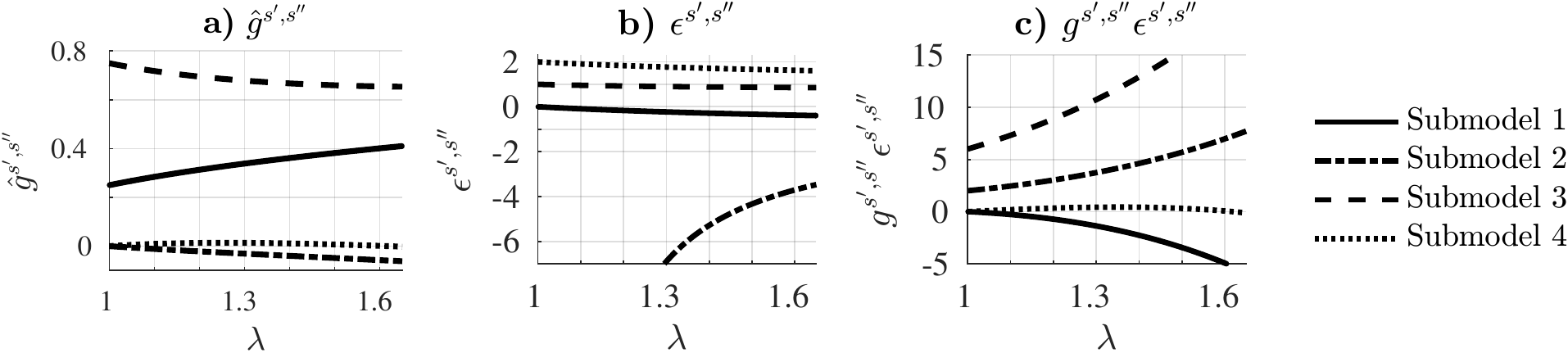}
\caption[]{The interaction coefficients. \label{fig:HNSE_shellmodel_params}}
\end{figure*}
\section{The new shell model}
\label{sec:NSSM}

The new model is obtained from the helically decomposed NSE \eqnref{eqn:Helical_NSE} in \suppmaterial \ref{sec:appendix_deriv} 
by defining complex velocity components $u_n^\Sk \equiv u^\Sk(k_n)$ ($\Sk=\pm$) on a exponentially thinned set of wave vector magnitudes $k_n = k_0\lambda^n$ for $n = 1,2, \cdots, N$.  
Within this discretised wave space, triadic interactions are permitted only between waves fulfilling the triangle inequality $\pqveccond$ (the shell model equivalent of the NSE constraint $\VK+\VP+\VQ=0$) and $\pqcond$ (a model constraint ensuring waves do not self-interact (unilateral triangles) such that Liouville's theorem is fulfilled as for the inviscid part of the NSE).

In the limit of nearest-neighbour interactions the new model is 
\begin{align}
\NSESMLocal + f^s_n, \label{eqn:NSSM_compact} 
\end{align}
where $\lambda$ and $k_0$ are the only free parameters and $\ush{n}{\Sk}{1}$ is the complex conjugate of $u_n^\Sk$.
Forcing and viscous dissipation at the $n$-th helical shell of sign $\Sk$ are $f_n^\Sk$ and $(\Disspdef)\ush{n}{\Sk}{0} \equiv \Dissp\ush{n}{\Sk}{0}$ respectively, $\DisspdefLS\ush{n}{\Sk}{0}$ being a large-scale drag added to remove any potential build-up of energy at large scales.
The summation over $\Sp$ and $\Sq$ is the weighted sum over the four \submodels
: for a given sign of $\Sk$ the dummy variables may take values of $\lbrace \Sp,\Sq\rbrace = \lbrace +,+\rbrace,\lbrace +,-\rbrace,\lbrace -,+\rbrace,\lbrace -,-\rbrace$, each pair representing one of the \submodels, yielding three helical interactions (per shell) per \submodel (figure \ref{fig:helicalcouplings}). 
The \submodelweights $\Wglocal$ and modal interaction coefficients $\Wepslocal$ and $\Wxilocal$ are given by (derived in \suppmaterial \ref{sec:appendix_deriv})
\begin{alignat}{3}
\Wglocal(\lambda)   &= \Wgdeflocal \label{eqn:weight_interact}\\
\Wepslocal(\lambda) &= \Wepsdeflocal \label{eqn:weight_modalcoef}  \\
\Wxilocal(\lambda)  &= \Wxideflocal \label{eqn:weight_modalcoef_xi} 
.
\end{alignat}
A note on the index notation used: The two helical sign indices $\Sp$ and $\Sq$ in $\Wglocal$, $\Wepslocal$ and $\Wxilocal$ are separated by a comma in order to distinguish them from the single index used for modal amplitudes (e.g. $\ush{n}{\Sk\Sp}{}$) in which helical signs are multiplied together to produce a resulting helical sign.

\citet{bib:waleffe1992nature} hypothesised three classes of triad interactions based on the average energy transfer directions using linear stability: 
1) a class of forward interactions consisting of triads with opposite helical signs of the two largest wave modes, 
2) a class of reverse interactions consisting of triads where all three helical signs are equal, and
3) a class of mixed forward/backward--behaving interactions (depending on triad shape) consisting of triads where the signs of the two largest wave modes are equal but unequal to the smallest. 
Because the product of helical signs of the two largest modes in all three non-linear terms of \eqnref{eqn:NSSM_compact} is $\Sp\Sq$ each \submodel belongs to one of the three above classifications.
Thus two of the \submodels are of the forward class ($\Sp\Sq=-1$, figure \ref{fig:helicalcouplings}.a,c), one of the reverse class ($\Sp\Sq=+1$ where $\Sk=\Sp=\Sq$, figure \ref{fig:helicalcouplings}.d), and one mixed type ($\Sp\Sq=+1$, figure \ref{fig:helicalcouplings}.b). 

\subsection{The interaction coefficients}
\label{sec:weights}
 
The interaction coefficients $\Wglocal$, $\Wepslocal$ and $\Wxilocal$ depend on the shell model spacing parameter $\lambda$, which indicates the triad geometry. For $\lambda\rightarrow 1$ triangles become equilateral while for $\lambda \rightarrow (1+\sqrt{5})/2$ (golden ratio) they collapse to a line.
The interaction coefficients $\Wglocal$ and $\Wepslocal$ are both plotted as functions of $\lambda$ in figure \ref{fig:HNSE_shellmodel_params}.
Panel \ref{fig:HNSE_shellmodel_params}.a shows the normalised \submodelweights $\Wgfuncnorm{}{\Sp,\Sq} = \Wglocal/\sum_{\Sp,\Sq}\Wglocal$.
Since two of the \submodels belong to the forward class their associated weights ($\Wgfunc{}{+,-}$ and $\Wgfunc{}{-,+}$) are expected to be largest because three-dimensional turbulence exhibits an average down-scale dominated energy cascade.
This is indeed found to be the case. 
The modal interaction coefficients $\Wepslocal$ are plotted in figure \ref{fig:HNSE_shellmodel_params}.b, indicating both forward $\Sp\Sq=-1$ \submodels have $\vert \Wepslocal \vert < 1$ whereas both the $\Sp\Sq=+1$ \submodels have $\vert \Wepslocal \vert > 1$. 
This is appealing because the structure of the new model \eqnref{eqn:NSSM_compact} and functional forms of $\Wepslocal$ \eqnref{eqn:weight_modalcoef} are somewhat similar to the helically decomposed GOY and Sabra counterparts. In GOY and Sabra models it is well known the limit $\Wepslocal = 1$ marks the transition between a two-dimensional behaviour of the energy cascade ($1<\Wepslocal <2$) and three-dimensional behaviour ($\Wepslocal<1$). Thus, the values of $\Wepslocal$ seem to support the expected forward/backward behaviour based on the $\Sp\Sq$ product. 

\subsection{Invariants and fluxes}
\label{sec:invars_and_fluxes}
\gdef\LLgenexpEpseudo{\genexpEsymb}
\gdef\LLgenexpHpseudo{\genexpHsymb}


\gdef\OneSidecorrLocalSymb#1#2#3{\Delta_{n}^{#1,#2,#3}}
\gdef\OneSidecorrLocal{
\OneSidecorrLocalSymb{\pm}{\Sp}{\Sq} \equiv 2 k_{n-1}\Re[
\triplehelampsshort{\KIDX-1}{*}{+}{\KIDX}{*}{\Sp}{\KIDX+1}{}{\Sq} \pm \triplehelampsshort{\KIDX-1}{*}{-}{\KIDX}{*}{-\Sp}{\KIDX+1}{}{-\Sq}]
}

\gdef\EfluxeqnLocal{\Efluxreal{n} &=         \Delta_{n+1}^{-,\Sp,\Sq} + (1 -\Sp\Wepslocal )\Delta_{n}^{-,\Sp,\Sq}}
\gdef\HfluxeqnLocal{\Hfluxreal{n} &= k_{n} ( \Delta_{n+1}^{+,\Sp,\Sq} + (\genH{-1} -\Wepslocal )\Delta_{n}^{+,\Sp,\Sq} )}


\gdef\piloglambda{\pi\I/\log(\lambda)}
\gdef\piloglambda{\frac{\pi\I}{\log{\lambda}}}
\begin{table}[!t]
\centering
\begin{tabular}{lrrr@{}lr@{}l}
 \toprule
           & \multicolumn{1}{c}{$\;\genE{\LLgenexpEpseudo}$} & \multicolumn{1}{c}{$\;\genE{\LLgenexpHpseudo}$} & \multicolumn{2}{c}{$\LLgenexpEpseudo$} & \multicolumn{2}{c}{$\LLgenexpHpseudo$} \\ 
\midrule
\Submodel 1 & $-\lambda$                            & $-1$                           & $1$     & $\,+\, \piloglambda$  &   $0$   & $\,+\,\piloglambda$ \\[0.3em]
\Submodel 2 & $\frac{\lambda-1}{\lambda+1}\lambda$  & $-\frac{\lambda-1}{\lambda+1}$ & $-6.76$ & $\,+\,0\I$            & $-7.76$ & $\,+\,\piloglambda$ \\[0.3em] 
\Submodel 3 & $-\frac{\lambda+1}{\lambda-1}\lambda$ & $\frac{\lambda+1}{\lambda-1}$  & $8.76$  & $\,+\, \piloglambda$  & $7.76$  & $\,+\,0\I$ \\[0.3em] 
\Submodel 4 & $\lambda$                             & 1                              & $1$     &  $\,+\,0\I$           & $0$     & $\,+\,0\I$  \\[0.3em] 
\bottomrule
\end{tabular} 
\caption{$\genE{\LLgenexpEpseudo}$ and $\genH{\LLgenexpHpseudo}$ solutions to \eqnref{eqn:NSSM_E_gen_eqn_LOCAL} and \eqnref{eqn:NSSM_H_gen_eqn_LOCAL} and the corresponding exponents for $\lambda=1.3$. \label{table:generatorset}}
\end{table}

Similarly to other shell models the non-linear terms in \eqref{eqn:NSSM_compact} conserve both energy and helicity. 
In the limit of nearest-neighbour interactions, however, each \submodel conserves two additional quadratic quantities 
\gdef\KIDXs{n}
\begin{alignat}{3}
\Einvarpseudo &= \sum_{n=1}^N  k_n^{\LLgenexpEpseudo} \modalcontr{+} \label{eqn:PSEUDO_INVARS_E} 
\\
\Hinvarpseudo &= \sum_{n=1}^N  k_n^{\LLgenexpHpseudo} \modalcontr{-}, \label{eqn:PSEUDO_INVARS_H} 
\end{alignat}
hereafter referred to as pseudo-energy and pseudo-helicity, both arising as a consequence of the way the spectral space is discretised in the shell model. The  exponents $\LLgenexpEpseudo$ and $\LLgenexpHpseudo$ are specific to each \submodel and are determined by calculating the rate of change of $\Einvarpseudo$ and $\Hinvarpseudo$ using \eqnref{eqn:NSSM_compact}. 
Doing so, one finds they respectively are constrained by (\suppmaterial \ref{sec:NSSM_props})
\begin{alignat}{9}
0 &= \Egeneqn{1}{2}{\LLgenexpEpseudo} 
\label{eqn:NSSM_E_gen_eqn_LOCAL}
\\
0 &= \Hgeneqn{1}{2}{\LLgenexpHpseudo} .
\label{eqn:NSSM_H_gen_eqn_LOCAL}
\end{alignat}
Thus, in addition to energy ($\LLgenexpEpseudo=0$) and helicity ($\LLgenexpHpseudo=1$) each \submodel conserves one pseudo-energy and one pseudo-helicity quantity with exponents $\LLgenexpEpseudo$ and $\LLgenexpHpseudo$ given by the relations in table \ref{table:generatorset}.

A detailed calculation of possible conserved quadratic quantities is presented in \suppmaterial \ref{sec:NSSM_props} where also non-nearest neighbour triads are considered. The existence of globally conserved (across all triad interactions) pseudo-invariants within each \submodels can potentially influence the behaviour of \submodel. However, because pseudo-invariants are not shared among \submodels only energy and helicity are globally conserved when mixing \submodels, which is similar to the NSE.

\gdef\KIDXs{m}
Non-linear spectral fluxes of energy and helicity through the $n$-th shell are given as the transfers from all wave numbers less than $k_n$ to wave numbers larger than, that is $\Efluxreal{n} = d_t\sum_{m=1}^n \modalcontr{+}$ and $\Hfluxreal{n} = d_t\sum_{m=1}^n k_m\modalcontr{-}$. Following the calculations through yields for a single \submodel (see \suppmaterial \ref{sec:NSSM_fluxes} for a generalised calculation)
\gdef\KIDXs{n}
\begin{alignat}{5}
\EfluxeqnLocal \label{eqn:NSSM_E_flux}\\
\HfluxeqnLocal, \label{eqn:NSSM_H_flux}
\end{alignat}
where the correlators are defined as $\OneSidecorrLocal$. 
For the \coupledmodel the corresponding expressions are merely weighted sums of \eqnref{eqn:NSSM_E_flux} and \eqnref{eqn:NSSM_H_flux} using weights $\Wglocal$.

\section{Numerical results}
\label{sec:numresults}

\gdef\forcingshellnr{18}
\gdef\modelnu{10^{-8}}
\gdef\modelsteps{5\times 10^{11}}
\gdef\modelsamplesteps{5\times 10^5}
\gdef\modelshells{53}
\gdef\frctypeboth{$f_{\forcingshellnr}^{+}, f_{\forcingshellnr}^{\pm}$ = const}

\begin{table}[!b]
\centering
\begin{tabular}{lccccccccc}
 \toprule
Experiment & $\Wgfuncnorm{}{-,+}$ & $\Wgfuncnorm{}{-,-}$ & $\Wgfuncnorm{}{+,-}$ & $\Wgfuncnorm{}{+,+}$ & $\Wepsfunc{}{\Sp,\Sq}$ & $\Wxifunc{}{\Sp,\Sq}$ & $\nu$  \\ 
\midrule
Coupled          & $0.34$ & $-0.03$ & $0.68$ & $0.01$ &   &  & $10^{-8\phantom{0}}$\\
\Submodel 1      & 1 & 0 & 0 & 0 & $-0.23$ & $-0.76$ & $10^{-7\phantom{0}}$\\
\Submodel 2      & 0 & 1 & 0 & 0 & $-6.89$ & $-5.89$ & $10^{-7\phantom{0}}$\\
\Submodel 3      & 0 & 0 & 1 & 0 &  $0.89$ &  $0.10$ & $10^{-8\phantom{0}}$\\
\Submodel 4      & 0 & 0 & 0 & 1 &  $1.76$ &  $0.76$ & $10^{-10}$\\
\bottomrule
\end{tabular}
\caption{Model configurations used in simulations. \label{table:modelconfigs}}
\end{table}

In order to investigate the model behaviour a set of simulations were conducted with and without input of helicity.
Following \cite{bib:biferale2012inverse,bib:biferale2013split,bib:sahoo2015} we performed "spectral surgery" in the sense that the behaviour of each \submodel was investigated separately in addition to a fully coupled model which includes all four \submodels. 
In total five distinct model configurations where integrated using $N=\modelshells$ shells and $\lambda=1.3$, the details of which are listed in table \ref{table:modelconfigs}. 
A fourth-order Runge-Kutta integration scheme was applied in all simulations using $dt=10^{-7}$ together with the forcing $f^\pm_{n_f} = (1+\I)/\ush{n_f}{\pm}{1}$ applied to shell $n_f = \forcingshellnr$, giving a constant input of energy $\meanenergy$ (not to be confused with the model parameter $\Wepslocal$), and helicity $\meanhelicity$. 
Due to the \submodel-dependant scaling of inertial ranges, the viscosity $\nu$ was chosen separately for each model configuration to ensure that dissipation occurs at the end of the resolved wave space, thereby producing the longest possible inertial ranges. The large-scale viscosity, however, was fixed at $\nu_L = 10^{2}$.
For each of the the five configurations two forcing scenarios were employed: one in which only the positive 18th helical shell is forced such that $\meanenergy = 2$, hereafter referred to as the $\meanhelicity \neq 0$ (helical) simulations (where $\meanhelicity  = k_{18} \meanenergy$), and one in which both 18th helical shells are forced such that $\meanenergy = 4$, hereafter referred to as the $\meanhelicity = 0$ (non-helical) simulations. 
All realisations are $10^{11}$ time-steps long and are initialised using the velocity profile $\ush{n}{\pm}{}\sim k_n^{-1/3}$. 
A spin-up of $10^{10}$ time-steps was performed to eliminate transients from the statistics. 
\gdef\modelfourcorrsymb#1{\Delta_{#1}^{\star}}
\gdef\correlatorratio{\langle\modelfourcorrsymb{n+1}\rangle/\langle\modelfourcorrsymb{n}\rangle}
\gdef\fluxratio{\langle \Efluxreal{n} \rangle/\langle\Hfluxreal{n} \rangle}
\gdef\funccorrrel{d_{n}}
\gdef\funccorrrellong{d_{n}(\fluxratio)}
\begin{figure*}[!t]
\centering
\includegraphics[scale=0.73]{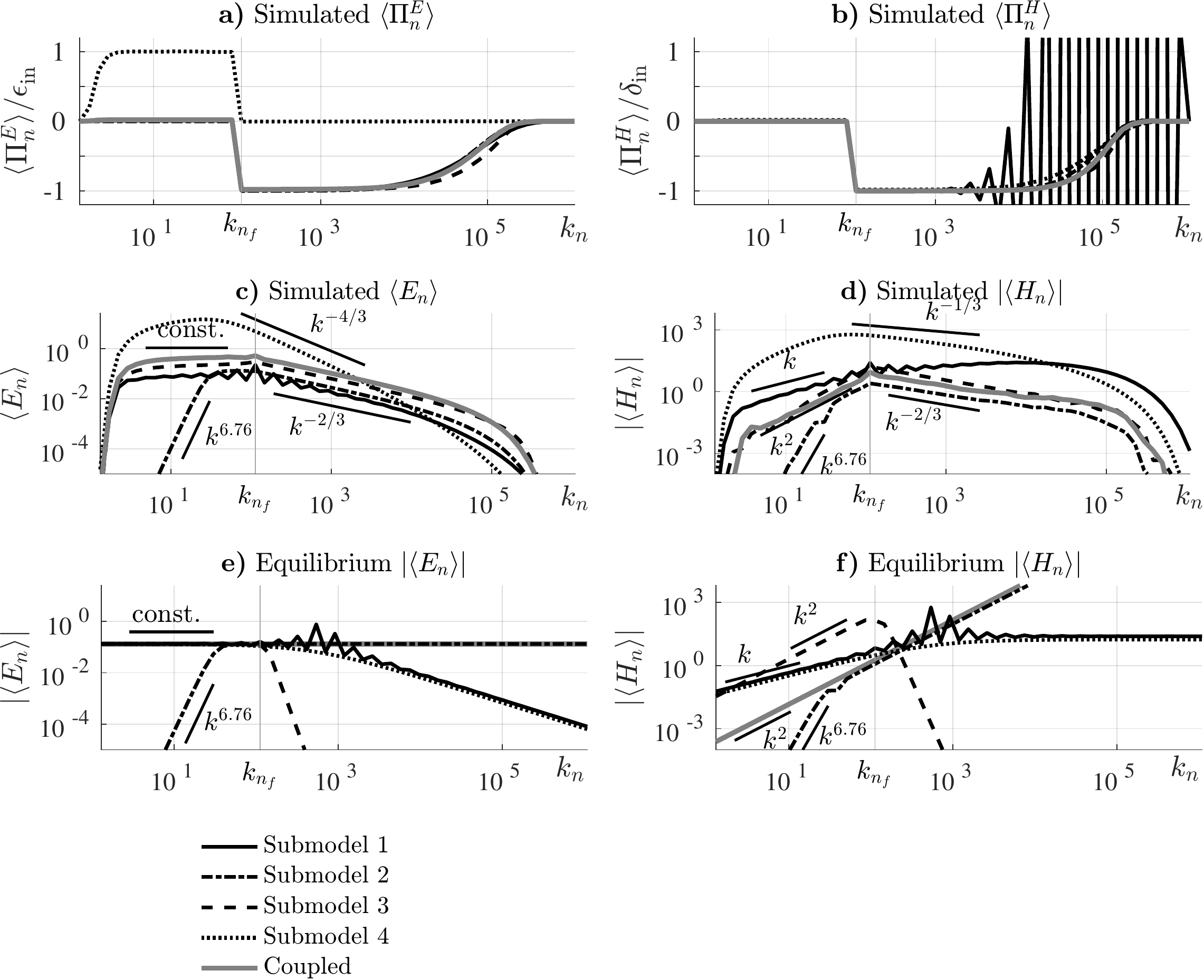} 
\caption{Energy and helicty spectra and fluxes (a)-(d), and predicted equlibrium spectra (e)-(f). \label{fig:results}}
\end{figure*}
\gdef\panelE{c\xspace}
\gdef\panelH{d\xspace}
\gdef\panelPiE{a\xspace}
\gdef\panelPiH{b\xspace}
\gdef\panelEeq{e\xspace}
\gdef\panelHeq{f\xspace}

Figures \ref{fig:results}.\panelE,\panelPiE show the energy spectra and fluxes, respectively, of the non-helically forced simulations ($\meanhelicity = 0$). The helically forced simulations ($\meanhelicity \neq 0$) are similar but not shown for clarity. The results show \submodels 1-3 and the \coupledmodel all exhibit expected K41 $\langle E_n\rangle \sim k_n^{-2/3}$ scaling energy spectra associated with a forward energy cascade for wave numbers $k_n>k_{n_f}$ (elaborated on below). For wave numbers $k_n<k_{n_f}$ (large scales) the three \submodels have distinctly different spectra. 
\Submodel 4, however, is different from \submodels 1-3 in the entire spectrum. This \submodel is found to transfer energy up-scale but does not exhibit any scaling of the energy spectrum for $k_n<k_{n_f}$. The scaling for wave numbers $k_n>k_{n_f}$ is approximately $\langle E_n\rangle \sim k_n^{-4/3}$ 
corresponding to a forward cascade of (positive) helicity.

Figures \ref{fig:results}.\panelH,\panelPiH show the helicity spectra and fluxes, respectively, of the helically forced simulations ($\meanhelicity \neq 0$). The results show \submodels 2, 3 and the \coupledmodel all exhibit $\sim k_n^{-2/3}$ scaling helicity spectra for $k_n>k_{n_f}$, which are accompanied by down-scale cascades of helicity. \Submodel 1 also exhibits a down-scale cascade of helicity but with a positive scaling helicity spectrum given simply by $\langle H_n \rangle \sim k_n\langle E_n\rangle$ (similarly to \submodel 4). The helicity spectrum of \submodel 4 does, similarly to the energy spectrum, not indicate any fixed scaling behaviour except for $k_{n_f} \leq k_n \lesssim 10^3$, but like the other \submodels also transfers helicity down-scale.

\gdef\Epart{\tempE + \temppseudoE k_n^{\LLgenexpEpseudo}}
\gdef\Hpart{   \tempH k_n + \temppseudoH k_n^{\LLgenexpHpseudo}}
\gdef\Hpartneg{\tempH k_n - \temppseudoH k_n^{\LLgenexpHpseudo}}
\gdef\EpartNorm#1{1+#1^\LLgenexpEpseudo\temppseudoE/\tempE}
\gdef\HpartNorm#1{#1+#1^\LLgenexpHpseudo\temppseudoH/\tempH}
\gdef\nf{n_{f}}
\gdef\kf{k_{\nf}}
\gdef\Eppower#1{^{#1\LLgenexpEpseudo}}
\gdef\Hppower#1{^{#1\LLgenexpHpseudo}}
\gdef\kfE#1{\kf\Eppower{#1}}
\gdef\kfH#1{\kf\Hppower{#1}}
\gdef\nEr{n_{\Einvarreal}}
\gdef\nEp{n_{\Einvarpseudo}}
\gdef\nHr{n_{\Hinvarreal}}
\gdef\nHp{n_{\Hinvarpseudo}}
\gdef\kEr{k_{\nEr}}
\gdef\kEp{k_{\nEp}}
\gdef\kHr{k_{\nHr}}
\gdef\kHp{k_{\nHp}}
\gdef\Dop{D}
\gdef\DEr{\Dop_{\nEr}}
\gdef\DEp{\Dop_{\nEp}}
\gdef\DHr{\Dop_{\nHr}}
\gdef\DHp{\Dop_{\nHp}}

In the following we differentiate between the parts of the simulated energy spectra in which flow invariants equipartition among the shells from those parts in which invariants cascade \citep{bib:ditlevsenmogensen1996}.
Using the equipartition theorem, a conservative system with quadratic invariants, in this case $\Einvarreal$, $\Einvarpseudo$, $\Hinvarreal$ and $\Hinvarpseudo$, will on average distribute the conserved quantities equally between the degrees of freedom in the system \citep{bib:robert1980two}. 
Thus, \submodel partition functions take the form $Z = \int \exp[-\sum_n ((\Epart+\Hpart)\USHvert{n}{+} + (\Epart-\Hpartneg)\USHvert{n}{-} )] 
\,\Pi_i d\USH{i}{+} d\USH{i}{-}$
where 
$\tempE,\temppseudoE,\tempH$, and $\temppseudoH$ are the inverse $\Einvarreal$, $\Einvarpseudo$, $\Hinvarreal$ and $\Hinvarpseudo$ temperatures respectively. 
Using the partition function the equilibrated energy and helicity spectra are easily calculated, giving 
\begin{align}
\langle \Einvarreal_n \rangle &= \frac{\Epart}{(\Epart)^2 - (\Hpart)^2} 
\approx \frac{1}{\Epart}
\label{eqn:regular_eqpart_E1}\\ 
\langle \Hinvarreal_n \rangle &= \frac{k_n(\Hpart)}{(\Epart)^2 - (\Hpart)^2}
\approx \frac{k_n(\Hpart)}{(\Epart)^2}
\label{eqn:regular_eqpart_H1}
\end{align}
where $\Epart\gg\Hpart$ has been used by noting that the energy spectra of the helical ($\meanhelicity \neq 0$) and non-helical ($\meanhelicity = 0$) simulations are similar (not shown).
The inverse temperatures may be constrained by using $\meanhelicity = \kf \meanenergy$,
$\meanenergypseudo = \kfE{} \meanenergy$ and $\meanhelicitypseudo = \kfH{} \meanenergy$, where $\nf$ is the forcing shell. 
Let the approximate (Kolmogorov) scales where invariants are dissipated be defined by the shell numbers $\nEr, \nEp, \nHr$ and $\nHp$, for $\Einvarreal$, $\Einvarpseudo$, $\Hinvarreal$ and $\Hinvarpseudo$ respectively. Conservation of the invariants then gives expressions for their approximate averaged dissipation
\begin{align}
\meanenergy &\approx \DEr\meanErspec{\nEr}, \quad 
\meanenergypseudo \approx \DEp\meanEpspec{\nEp}, \quad
\\
\meanhelicity &\approx \DHr\meanHrspec{\nHr} 
,\quad
\meanhelicitypseudo \approx \DHp\meanHpspec{\nHp} 
.
\end{align}
\begin{table}[!t]
\centering
\begin{tabular}{lrrrrrrrr}
 \toprule
                 & $\kEr$ & $\kEp$ & $\kHr$ & $\kHp$  \\ 
\midrule
Coupled          & \SI{5e3}{} &              & \SI{1e3}{} &     \\
\Submodel 1      & \SI{5e3}{} & $\kHr$       & \SI{5.6e2}{} & $\kEr$\\
\Submodel 2      & \SI{5e3}{} & \SI{5e0}{}   & \SI{1e3}{} & \SI{7e0}{}\\
\Submodel 3      & \SI{5e3}{} & \SI{1.1e2}{} & \SI{1e4}{} & \SI{1.3e2}{}\\
\Submodel 4      & \SI{4e0}{} & $\kHr$       & \SI{1e3}{} & $\kEr$\\
\bottomrule
\end{tabular}
\caption{Approximate dissipation scales.\label{table:temperature_dependencies}}
\end{table}
Combining the above expressions all temperatures are related to $\tempE$ by
\begin{align}
\tempE/\temppseudoE &= \frac{\DEr\kEp\Eppower{-} - \DEp\kfE{-}}{\DEp\kfE{-}\kEp\Eppower{} - \DEr} \label{eqn:temp_rel_aa}\\
\tempH/\tempE       &= \frac{\DEr\kf(\EpartNorm{\kHr})^2}{\DHr\kHr(\EpartNorm{\kEr})(\HpartNorm{\kHr})} \label{eqn:temp_rel_ba}\\
\temppseudoH/\tempH &= \dfrac{\splitdfrac{(\EpartNorm{\kHp})\DHr\kf^{\LLgenexpHpseudo-1}\kHr^2}{ - (\EpartNorm{\kHr})\DHp\kHp^{\LLgenexpHpseudo+1}}}{\splitdfrac{(\EpartNorm{\kHr})\DHp\kHp^{2\LLgenexpHpseudo}}{ - (\EpartNorm{\kHp})\DHr\kf^{\LLgenexpHpseudo-1}\kHr^{\LLgenexpHpseudo+1}}}\label{eqn:temp_rel_bb}
.
\end{align} 
Inserting \eqnref{eqn:temp_rel_aa}--\eqnref{eqn:temp_rel_bb} into \eqnref{eqn:regular_eqpart_E1} and \eqnref{eqn:regular_eqpart_H1} figure \ref{fig:results}.\panelEeq,\panelHeq shows each \submodel equilibrium spectra with $\tempE = 30$ (for offsets comparable to figures \ref{fig:results}.\panelE,\panelH) and $\nf = 18$ (as in simulations) using dissipation scales obtained from best fits to the simulated spectra in figures \ref{fig:results}.\panelPiE-\panelH and corresponding pseudo-invariant plots (not shown), all of which are listed in table \ref{table:temperature_dependencies}. 

Comparing the simulated $\meanErspec{n}$ and $\meanHrspec{n}$ spectra of \submodels 1-3 and the \coupledmodel with the equilibrium spectra one finds that these agree well, suggesting equipartitioning of energy ($E_n$) and pseudo-energy ($k_{n}^{\LLgenexpEpseudo}E_n$) for $k_n<k_{n_f}$. 
The slightly weak positive scaling of $\meanErspec{n}$ in \submodels 1 and 3 and the \coupledmodel are due to an insufficiently short spectral range connecting the forcing scale with the large-scale sink. This is evident from identical simulations using a smaller-scale forcing ($n_f = 36$) in which $\meanErspec{n}$ are flat for $k_n<k_{n_f}$ (not shown).
The simulated $\meanHrspec{n}$ spectra of \submodels 1-3 also match the expected equilibrium spectra of figure \ref{fig:results}.\panelHeq for $k_n<k_{n_f}$, which remarkably even captures the small dip exhibited by \submodel 2.

Before moving on to \submodel 4 consider the scaling behaviour for wave numbers $k_n>k_{n_f}$ of \submodels 1-3 and the \coupledmodel. 
There, the non-linear energy flux is constant which is fulfilled if the correlators scale as $\OneSidecorrLocalSymb{-}{\Sp}{\Sq} \sim \text{const.}$, implying velocity components scale as $\USH{n}{\pm} \sim k_n^{-1/3}$. Thus, one would expect $\meanErspec{n} \sim k_n^{-2/3}$ which is indeed found to be the case.
The energy and helicity fluxes (figures \ref{fig:results}.\panelPiE,\panelPiH) indicate dual down-scale cascades of energy and helicity in \submodels 1-3 and the \coupledmodel. \citet{bib:brissaud1973helicity} envisaged that such dual down-scale cascades would manifest themselves by the helicity spectrum scaling linearly with the energy spectrum, i.e. $\meanErspec{n}\sim \meanHrspec{n} \sim k_n^{-2/3}$, which is also supported by the present study (figure \ref{fig:results}.\panelH).

\begin{figure}[b]
\centering
\includegraphics[scale=0.60]{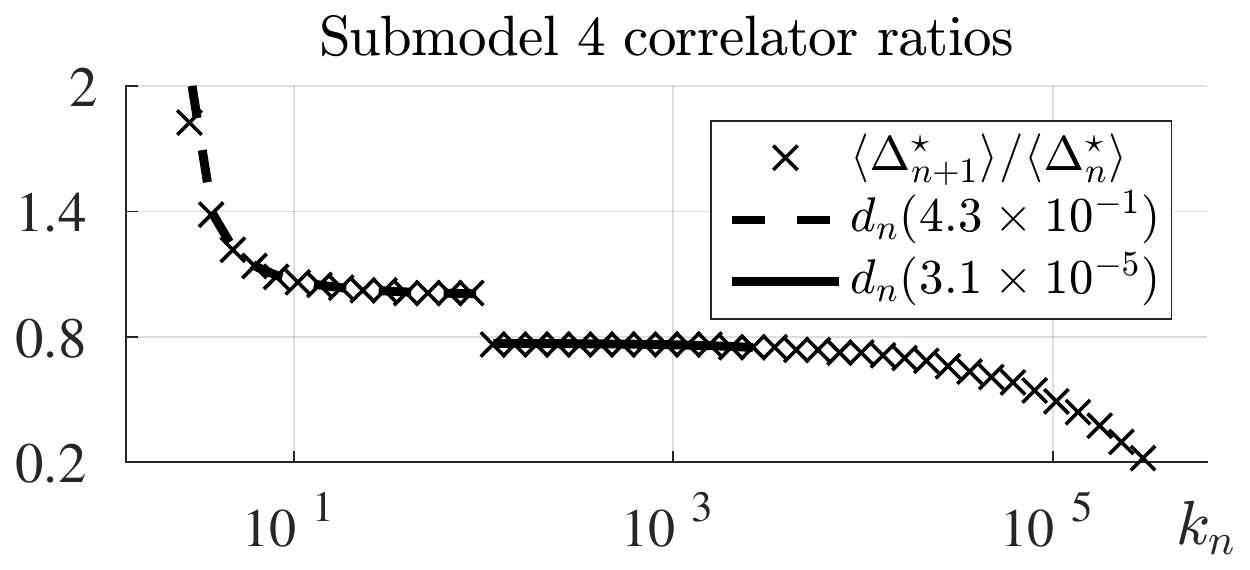} 
\caption{\Submodel 4 correlator ratios.\label{fig:submodel4_corr_scalings}}
\end{figure}

\begin{figure*}[!t]
\centering
\includegraphics[scale=0.89]{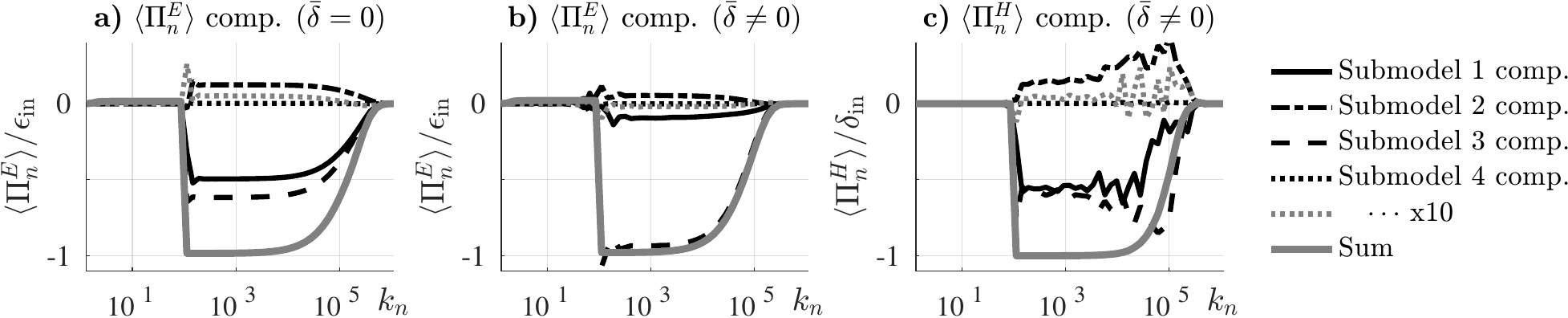} 
\caption{\Submodel components of the \coupledmodel energy and helicity fluxes. \label{fig:submodelcontr}}
\end{figure*}

The energy and helicity spectra of the fourth \submodel are not found to resemble the equilibrium spectra, suggesting that equipartitioning of flow invariants are not responsible for the shape of the spectra.
Following the above K41 scaling argument one might expect the energy spectrum to scale as $\sim k_n^{-2/3}$ for $k_n<k_{n_f}$ due to the energy cascade and $\sim k_n^{-4/3}$ for $k_n>k_{n_f}$ due to the helicity cascade (by a similar argument), but this is clearly not the case either. 
The failure of the K41 argument may be understood from the specific ratios $\fluxratio$ in the two inertial ranges of the flow, which allow the correlators to be scale dependent while simultaneously supporting constant energy and helicity fluxes. 
In \submodel 4 helical modes of opposite signs do not interact, thus if there is no pumping of a specific sign of helicity all modes of that sign will decay.
In this case the correlators reduce to $\OneSidecorrLocalSymb{+}{+}{+} = \OneSidecorrLocalSymb{-}{+}{+} = 2 k_{n-1}\Re[\triplehelampsshort{\KIDX-1}{*}{+}{\KIDX}{*}{+}{\KIDX+1}{}{+}] \equiv \modelfourcorrsymb{n}$. Calculating the ratio $\fluxratio$ by inserting $\modelfourcorrsymb{n}$ into \eqnref{eqn:NSSM_E_flux} and \eqnref{eqn:NSSM_H_flux}, one finds the exact relation
\begin{align}
\frac{\langle\modelfourcorrsymb{n+1}\rangle}{\langle\modelfourcorrsymb{n}\rangle} = 
\frac{\Wepsfunc{}{+,+} - k_n\fluxratio - 1}{1-k_n\fluxratio} 
\equiv \funccorrrellong, \label{eqn:model4_corr_scaling}
\end{align}
which may be scale sensitive depending on $\fluxratio$. The simulated ratios are found to be $\fluxratio = 4.3\times 10^{-1}$ in the inertial range $k_n<k_{n_f}$ (shell 5-17 average) and $\fluxratio = 3.1\times 10^{-5}$ in the intertial range $k_n>k_{n_f}$ (shell 19-31 average). 
Using these ratios, figure \ref{fig:submodel4_corr_scalings} shows the simulated $\correlatorratio$ values compared to the anticipated $\funccorrrel$ forms (plotted only in their valid ranges where flux ratios are constant). The correlators clearly exhibit scale dependence for $k_n<k_{n_f}$ following $\funccorrrel$, thus suggesting the K41 argument leading to $\meanErspec{n} \sim k_n^{-2/3}$ is not necessarily valid. 
For larger wave numbers, however, $\correlatorratio$ seems constant across many shells. By closer inspection $\funccorrrel$ does not start diverging until $k_n \approx 10^3$, suggesting $\meanErspec{n} \sim k_n^{-4/3}$ might be expected up until that scale, which is indeed the case (figure \ref{fig:results}.\panelE). 
From thereon $\funccorrrel$ changes slowly with $k_n$, suggesting a slow change in the scaling of $\langle E_n \rangle$ compared to $k_n < k_{n_f}$, which is also found to be the case.

Finally, figure \ref{fig:submodelcontr} shows the \coupledmodel fluxes $\langle\Efluxreal{n}\rangle$ and $\langle\Hfluxreal{n}\rangle$ split into their four \submodel contributions/components, given by \eqnref{eqn:NSSM_E_flux} and \eqnref{eqn:NSSM_H_flux} multiplied by their \submodelweights \eqnref{eqn:weight_interact}. Figure \ref{fig:submodelcontr}.a shows the forward energy cascade in non-helical turbulence is predominantly carried by \submodel 1 and 3 interactions, whereas  \submodels 2 and 4 both contribute with relatively small up-scale cascades, the former being more than an order of manitude larger than the latter. In helical turbulence, however, the forward energy cascade is carried almost entirely by \submodel 3 interactions (figure \ref{fig:submodelcontr}.b), whereas the forward helicity cascade is dominated equally by \submodel 1 and 3 interactions while \submodel 2 contributes with a small reverse component (figure \ref{fig:submodelcontr}.c). 
Thus, two important results emerge from the coupled simulations: 1) the behaviour of \submodel 2 is flipped in a coupled configuration, sending energy and helicity up-scale instead of down-scale when considered alone, and 
2) the set of helical interactions dominating energy cascade dynamics in non-helical turbulence are different from those in helical turbulence.

\section{Comparison to other shell models}
\label{sec:discussion}

The shell model introduced here is obtained from the helical decomposition of the NSE.
It is remarkable the three helical interactions (per shell) of each \submodel are identical to those of helically decomposed GOY and Sabra \submodels apart from the interaction coefficients \citep{bib:biferale1995role,bib:benzi1996helical,bib:biferale1998helicityA,bib:biferale1998helicityB,bib:DePietro2015arXiv150806390D}. In fact, the individual \submodel equations \eqnref{eqn:NSSM_compact} are quite similar to the helical Sabra \submodels, suggesting the new coupling weights could be used for the Sabra model as well.
\citet{bib:benzi1996helical} implemented the four helical \submodels in a GOY model. Interestingly, the (absolute) values of $\Wepslocal$ indicate the new model similarly to the GOY model consists of two \submodels (1 and 4) with canonical 2D and 3D $\Wepslocal$-configurations, and one new 3D type (section \ref{sec:weights}).
The last \submodel (\submodel 2) was found by \citet{bib:benzi1996helical} to show signs of a reverse energy cascade, a property not shared by the new model in its nearest-neighbour limit.

Recent work by \citet{bib:DePietro2015arXiv150806390D} also numerically investigated the Sabra model equivalent of \submodel 2.
In their work, however, they found the energy spectrum to scale like $\sim k_n^{-0.28}$ for wave numbers $k_n<k_{n_f}$ as opposed to energy/pseudo-energy equipartitioning as found here. The discrepancy might be due to dissimilar shell spacings $\lambda$ being used, however conducting such further numerical experiments is beyond the scope of the present work.
Additionally, \citet{bib:DePietro2015arXiv150806390D} found the second nearest-neighbour set of interactions in \submodel 2, i.e. between shells such as $n,n+2,n+3$ ($p,q = 2,3$) as opposed to $n,n+1,n+2$ ($p,q=1,2$), produces a reverse energy cascade as opposed to a forward cascade in the nearest-neighbour case.
Using their methodology where $f \equiv q+(q-p)\Weps$ should predict a reverse energy cascade whenever $f > 0$ and a forward cascade whenever $f < 0$, one finds that only a small $\pqset$-subset is predicted to contribute reversely for a given $\lambda$ spacing. As an example the $\lambda = 1.3$ configuration used here allows for five triad geometries, none of which have $f>0$, whereas $\lambda=1.2$ allows for thirteen triads of which only two have $f>0$.
It would be interesting to investigate if this property is indeed shared by the new model, however, such simulations are also beyond the scope of the present work. Although this property might be shared by the new model too, this reverse behaviour is possibly suppressed in a multi-triad configuration if the number of reversely contributing triads is as small as the $f$-prediction would suggest.

\citet{bib:gilbert2002inverse} showed that a regular Sabra model in the 2D configuration $\Wepsfunc{}{+,+} > 1$, corresponding to \submodel 4 here, exhibits different energy scaling regimes depending on the value of $\Wepsfunc{}{+,+}$. Their works suggests that whenever $\Wepsfunc{}{+,+}/\lambda < 1+\lambda^{-2/3}$ the reverse energy flux regime should be accompanied by a proper K41 scaling energy spectrum, whereas above this critical value a quasiequilibrium energy spectrum should develop. Inserting $\Wepsfunc{}{+,+}$ from \eqnref{eqn:weight_modalcoef} one would thus always expect a K41 scaling to occur. However, present simulations can hardly be said to scale as $\sim k_n^{-2/3}$ nor be in quasiequilibrium for $k_n<k_{n_f}$. In order to further compare \submodel 4 with their work additional simulations were conducted using $\lambda = \lbrace 1.1,1.2,1.3,1.4,1.5,1.6,2.0\rbrace$ with $N = \lbrace 146,  76, 53,   41,  37,  34,  22\rbrace$ respectively (ensuring $k_N$ are roughly the same). In all cases energy spectra were found to behave as shown in figure \ref{fig:results}.\panelE (not shown), suggesting the work by \citet{bib:gilbert2002inverse} does not carry over to \submodel 4 of the new model.

\section{Discussion and summary}
\label{sec:conclusion}
The role of helicity in 3D turbulence was numerically investigated in the context of a new shell model obtained as a special case of the helically decomposed Navier--Stokes equation (NSE) \citep{bib:waleffe1992nature}.
Numerical experiments were performed of the four naturally occurring subsets of interactions (\submodels) in the limit of local triadic interactions, which share strong similarities with the four existing helically decomposed Sabra shell models.
In accordance with expectations, results show three of the four \submodels (\submodels 1-3) contribute with dual down-scale cascades of energy and helicity, whereas the last \submodel (\submodel 4), which renders both signs of helicity separately (inviscidly) conserved, transfers energy up-scale and helicity down-scale.
The behaviour of the \coupledmodel is found to be strongly dominated by dual forward (down-scale) cascades of energy and helicity, which matches expectations based on the magnitudes of the submodel coupling weights and the fact that real three-dimensional turbulence should be dominated by such dual forward transfers.
By applying both helical and non-helical mid-scale forcings it was found that the forward energy cascade in helical turbulence is carried almost entirely by \submodel 3 interactions whereas \submodel 1 and 3 contribute roughly equally in non-helical turbulence. 

In the \coupledmodel and the three dual-cascading \submodels (\submodels 1-3) flow invariants were found to equipartition in the range of scales $k_n<k_{n_f}$ ($k_{n_f}$ being the forcing scale), which was explained using the equipartition theorem with multiple conserved quadratic quantities.
The remaining \submodel (\submodel 4) exhibits a reverse energy cascade, but has a very small weight in comparison to the other \submodels in the full set of triadic interactions of the helically decomposed dynamics.
By investigating the scaling behaviour of the triple correlations used in energy and helicity flux calculations it was found these can not necessarily be assumed scale-independent within inertial ranges (as one would otherwise expect compared to the other \submodels), thus breaking the anticipated energy spectrum scaling within inertial ranges.

\ifx\DOFINAL\undefined
	\appendix
\section{The new shell model}
\label{sec:appendix_deriv}

\newcommand{\ABC}[5]{
	\ifthenelse{\equal{#1}{1}}{
   		\ifthenelse{\equal{#2}{+}\OR\equal{#2}{-}}{ \effsymnp{#2}{+\lambda^{#4}} }{ \effsymnp{+1}{#2}\lambda^{#4}}
   		\effsym{-1}{#3}\lambda^{#5} 	}{
	\ifthenelse{\equal{#1}{2}}{
		\ifthenelse{\equal{#2}{+}\OR\equal{#2}{-}}{ \effsymnp{#2}{+1} }{ \effsymnp{+1}{#2} } 
		\effsym{-1}{#3}\lambda^{#5} 	}{
	\ifthenelse{\equal{#1}{3}}{
		\ifthenelse{\equal{#2}{+}\OR\equal{#2}{-}}{ \effsymnp{#2}{+1} }{ \effsymnp{+1}{#2}} 
		\effsym{-1}{#3}\lambda^{#4} 	}{
	}}}
}
\newcommand{\ABCfrac}[3]{
	\ifthenelse{\equal{#1}{1}}{ 	(\ABC{#1}{#2}{#3}{p}{q})  	}{
	\ifthenelse{\equal{#1}{2}}{ 	\frac{\ABC{#1}{#2}{#3}{p}{q}}{\lambda^p} 	}{
	\ifthenelse{\equal{#1}{3}}{ 	\frac{\ABC{#1}{#2}{#3}{p}{q}}{\lambda^q}	}{
	}}}
}

\newcommand{\gfac}[9]{\triadgeom{#1}{#7}{#2}{#3}{#8}{#4}{#5}{#9}{#6}} 
\newcommand{\shapedef}[5]{\Lambda_{#1,#2,#3}^{#4,#5}}
\newcommand{\shapedefclean}[5]{\Lambda_{#1#2#3}^{#4#5}}
\newcommand{\shape}[4]{		
	\ifthenelse{\equal{#1}{1}}{ \shapedef{\effsymnp{+1}{#2}}{\effsymnp{-1}{#3}}{\effsymnp{+1}{#4}}{p}{q} }{ 
	\ifthenelse{\equal{#1}{2}}{ \shapedef{\effsymnp{+1}{#4}}{\effsymnp{-1}{#3}}{\effsymnp{+1}{#2}}{p}{q} }{ 
	\ifthenelse{\equal{#1}{3}}{ \shapedef{\effsymnp{-1}{#3}}{\effsymnp{+1}{#4}}{\effsymnp{-1}{#2}}{p}{q} }{ 
	}}}
} 

\newcommand{\triadclass}[6]{ 
	\ifthenelse{\equal{#6}{}}{\xdef\mycolor{Black}}{\xdef\mycolor{#6}} 
	\ifthenelse{\equal{#2}{3}}{\xdef\psign{-1}}{\xdef\psign{+1}} 
	\xdef\vecentry{\intcalcMul{1}{#2}}
	\xdef\lenentry{\intcalcMul{-1}{#2}}
	\ifthenelse{\equal{#1}{0}}{ 
		(#4\vp{\lenentry} - #5\vq{\lenentry})
		\;\;&& 
		\gfac{#4}{\vp{\vecentry}}{#5}{\vq{\vecentry}}{#3}{\vk{\vecentry}}{}{}{*}
		\;\;&& 
		\hamps{0}{#4}{\vp{\vecentry}}{#5}{\vq{\vecentry}} 
		}{
	\ifthenelse{\equal{#1}{1}}{ 
		(#4\vp{\lenentry} - #5\vq{\lenentry})
		\;\;
		\ifthenelse{\equal{#2}{1}}{ \gfac{#4}{\vp{\vecentry}}{-#5}{\vq{\vecentry}}{#3}{\vk{\vecentry}}{*}{*}{*} }{ 
		\ifthenelse{\equal{#2}{2}}{ \gfac{#3}{\vk{\vecentry}}{-#5}{\vq{\vecentry}}{#4}{\vp{\vecentry}}{*}{*}{*} }{ 
		\ifthenelse{\equal{#2}{3}}{ \gfac{-#5}{\vq{\vecentry}}{#3}{\vk{\vecentry}}{-#4}{\vp{\vecentry}}{*}{*}{*} }{ 
		}}}
		\;\;
		\ifthenelse{\equal{#2}{1}}{ \helampslong{#4}{*}{\vp{\vecentry}}{#5}{ }{\vq{\vecentry}} }{ 
		\ifthenelse{\equal{#2}{2}}{ \helampslong{#4}{*}{\vp{\vecentry}}{#5}{ }{\vq{\vecentry}} }{ 
		\ifthenelse{\equal{#2}{3}}{ \helampslong{#4}{}{\vp{\vecentry}}{#5}{ }{\vq{\vecentry}} }{ 
		}}}
		}{
	\ifthenelse{\equal{#1}{5}}{ 
		\ABCfrac{#2}{#4}{#5} 
		{\color{\mycolor} \shape{#2}{#4}{#5}{#3} }
		\hamps{\intcalcMul{-1}{#2}}{#4}{\protidx{#2}}{#5}{\qrotidx{#2}}
		}{  
	}}}
}

\newcommand{\condlow}{\vk{-1}<\vp{-1}<\vq{-1}}
\newcommand{\condmid}{\vp{-2}<\vk{-2}<\vq{-2}}
\newcommand{\condhigh}{\vp{-3}<\vq{-3}<\vk{-3}}
\gdef\colorA{black}
\gdef\colorB{black}
\gdef\colorC{black}
\gdef\colorD{black}

\def\wavespace{$\VK$-space\xspace}

\def\triadgrp{group\xspace} 
\def\triadgrpcap{Group\xspace}
\def\triadgrpplural{groups\xspace} 

In order to obtain the new model from the helically decomposed Navier--Stokes equation (NSE) \eqnref{eqn:Helical_NSE} it is necessary to impose two constraints: 1) assume spectral velocity components are independent of direction in \wavespace $\uhel{\Sk}{}{\VK} = \uhel{\Sk}{}{\VKLEN\VKHAT} = \uhel{\Sk}{}{\VKLEN}$,
and 2) reducing \wavespace to include only components which are increasingly spaced in magnitude according to the geometrical progression $k_n = k_0\lambda^n$ for $n = 1, 2, \cdots, N$.
Within this wave set only cross-scale triadic interactions are considered, i.e. triads in which all three wave components have different magnitudes, which is inspired by the structure of GOY and Sabra shell models.
Since only cross-magnitude interactions are considered, it is useful to split the triadic sum in the NSE \eqnref{eqn:Helical_NSE} into three separate sums, hereafter referred to as the three triad \emph{\triadgrpplural}, for which $\VK$ is the smallest ($\condlow$), middle ($\condmid$) and largest ($\condhigh$) wave number. 
Note that double primed vectors are chosen to be larger than single primed which leads to no loss of generality due to symmetry when interchanging the dummy waves $\VP\leftrightarrow\VQ$ (and $\Sp\leftrightarrow\Sq$).
Additionally, the vectorial condition $\vk{1}+\vp{1} + \vq{1} = 0$ on each triadic sum can be re-written by expressing the largest mode as a sum of the two smaller and absorbing the resulting negative signs into the terms of the sums using reality $\uf{}{-k} = \uu{}{*}{k}{}$ and the basis property $\hs{s}{}{-k} = \hs{-s}{}{k}$ \citep{bib:waleffe1992nature}. The vectorial conditions on each triadic sum thus become $\vk{1}+\vp{1} = \vq{1}$, $\vk{2}+\vp{2}=\vq{2}$ and $\vk{3}=\vp{3}+\vq{3}$ for \triadgrpplural 1--3 respectively) and the NSE \eqnref{eqn:Helical_NSE} then takes the form
\begin{widetext}
\begin{alignat}{4}
(\partial_t + \nu k^2) \uhel{\Sk}{}{\vk{1}} = -\frac{1}{4}\sum_{\Sp, \Sq}\Big[
\;\;\;& \sum_{\mathclap{\substack{\vk{1}+\vp{1} = \vq{1}\\ \mathrm{where}\; \condlow}}} \;\triadclass{1}{1}{\Sk}{\Sp}{\Sq}{}   \notag \\
-& \sum_{\mathclap{\substack{\vk{2}+\vp{2} =\vq{2}\\ \mathrm{where}\; \condmid}}} \;\triadclass{1}{2}{\Sk}{\Sp}{\Sq}{}   \notag \\
+& \sum_{\mathclap{\substack{\vk{3}=\vp{3}+\vq{3}\\ \mathrm{where}\; \condhigh}}} \;\triadclass{1}{3}{\Sk}{\Sp}{\Sq}{}  \Big] \label{eqn:HNSE_truncated1}
\end{alignat}
\end{widetext}
where the anti-symmetric property of $\gfac{\Sp}{\vect{k}\sprime}{\Sq}{\vect{k}\dprime}{\Sk}{\vect{k}}{*}{*}{*}$ has been used to re-arrange the order of basis components in a way which shall be useful later.
Consider now re-writing \eqnref{eqn:HNSE_truncated1} by:
\begin{table}[b]
\centering
\begin{tabular}{l@{}c@{}r@{\hskip 1em}l@{}c@{}l@{}c@{}r@{\hskip 1em}l@{}c@{}l@{}c@{}r}
\toprule
\multicolumn{3}{c}{\triadgrpcap 1} & \multicolumn{5}{c}{\triadgrpcap 2} & \multicolumn{5}{c}{\triadgrpcap 3} \\\midrule
$\vk{1}$&$\,=\,$& $\vkrotabs{1}\,\,$ &   $\vk{2}$&$\,\rightarrow\,$ & $\vkrotrel{2}$&$\,=\,$ & $\vkrotabs{2}\,$   &    $\vk{3}$ & $\,\rightarrow\,$&$\vkrotrel{3}$&$\,=\,$ & $\vkrotabs{3}$ \\[0.4em] 
$\vp{1}$&$\,=\,$& $\vprotabs{1}\,$   &   $\vp{2}$&$\,\rightarrow\,$ & $\vprotrel{2}$&$\,=\,$ & $\vprotabs{2}\,\,$ &    $\vp{3}$ & $\,\rightarrow\,$&$\vprotrel{3}$&$\,=\,$ & $\vprotabs{3}\,\,$ \\[0.4em] 
$\vq{1}$&$\,=\,$& $\vqrotabs{1}$     &   $\vq{2}$&$\,\rightarrow\,$ & $\vqrotrel{2}$&$\,=\,$ & $\vqrotabs{2}$     &    $\vq{3}$ & $\,\rightarrow\,$&$\vqrotrel{3}$&$\,=\,$ & $\vqrotabs{3}\,$ \\ 
\bottomrule
\end{tabular} 
\caption{Re-expressed wave vectors of rotated triads.\label{table:rot_wave_vectors}}
\end{table}
\begin{enumerate}
\item Reducing \wavespace to include only components with magnitudes given by $k_n = k_0\lambda^n$ and assuming direction independence of $\uhel{\pm}{}{\VK}$ in \wavespace.
Depending on $\lambda$ the triangle inequality constrains the possible choices of $n$ in $k_n$ which can be combined to construct triads. 
In the interest of generality, consider therefore the range of integers $p$ and $q$ which fulfil the triangle inequality $\pqveccond$. 
Crucially, given any $\lbrace\lambda,p,q\rbrace$-set the \wavespace reduction implies all possible triads have identical shapes independently of $n$. 
This motivates the use of direction independence in \wavespace because it allows rotating all triads of a given $\lbrace\lambda,p,q\rbrace$-set into a shared orientation in which they are simply $\lambda$-multiples each other. 
By rotating triads of \triadgrp 2 and 3 into the same orientation as \triadgrp 1 they can be re-expressed as given in table \ref{table:rot_wave_vectors}.
Inserting the rotated wave vectors into \eqnref{eqn:HNSE_truncated1} the three triadic sums can be made similar and re-joined into one sum with three terms. 
Because $\uhel{\pm}{}{\VK}$ is assumed independent of $\VK$'s direction it is unnecessary to take the joined sum over all possible directions in \wavespace since each contribution will be equal (given some fixed magnitudes $\VPLEN$ and $\VQLEN$).
Thus only one mode per magnitude really needs to be resolved, implying the sum is redundant and may be dropped.

Because the above analysis applies to any $\lbrace\lambda,p,q\rbrace$-set, the sum over triad geometries in the NSE is effectively reduced to a summation over integers $p$ and $q$ fulfilling the triangle inequality $\pqveccond$ (given some $\lambda$) and $\pqcond$.

\item Having assumed direction independence in \wavespace the rotational term (complex exponential) in the geometry term is assumed discardable. The geometry term can thus be written in the more compact notation
\begin{align}
\hslong{\Sp}{*}{\VP}\times\hslong{\Sq}{*}{\VQ} \cdot \hslong{\Sk}{*}{\VK}  
= \mybreak -\frac{Q(\VKLEN,\VPLEN,\VQLEN)}{2\VKLEN\VPLEN\VQLEN}\Sk\Sp\Sq(\Sk\VKLEN + \Sp\VPLEN + \Sq\VQLEN) 
\mybreak{}
\equiv \shapedefclean{\Sp,}{\Sq,}{\Sk}{}{}(\lambda^p,\lambda^q,1)  
\equiv \shapedef{\Sp}{\Sq}{\Sk}{p}{q}
\label{eqn:triad_shape_def}
\end{align}
where the scale-independent property has been used 
$\shapedefclean{\Sp,}{\Sq,}{\Sk}{}{}(\VPLEN,\VQLEN,\VKLEN) = \shapedefclean{\Sp,}{\Sq,}{\Sk}{}{}(\lambda^p,\lambda^q,1)$
and $Q(\VKLEN,\VPLEN,\VQLEN)=(2\VKLEN^2 \VPLEN^2 + 2\VPLEN^2 \VQLEN^2 + 2\VQLEN^2 \VKLEN^2 - \VKLEN^4 - \VPLEN^4 - \VQLEN^4)^{1/2}$ (see \citet{bib:waleffe1992nature} for details). 
\end{enumerate}
Following the two above steps through one finds that \eqnref{eqn:HNSE_truncated1} becomes
\begin{alignat}{4}
(d_t + \nu \VKLEN_n^2)& \mathrlap{\uhel{\Sk}{}{\VKLEN_n} 
= -\frac{1}{4}\VKLEN_n \sum_{\pqcond}\sum_{\Sp,\,\Sq} \,\Big[ }
\mybreak 
 &&\ABCfrac{1}{+\Sp}{+\Sq}  & \shape{1}{+\Sp}{+\Sq}{+\Sk} &&  \hamps{-1}{\Sp}{\protidx{1}}{\Sq}{\qrotidx{1}} \notag\\[-0.5em]
&&-\;\ABCfrac{2}{+\Sp}{+\Sq}  & \shape{2}{+\Sp}{+\Sq}{+\Sk} &&  \hamps{-2}{\Sp}{\protidx{2}}{\Sq}{\qrotidx{2}} \notag\\
&&+\;\ABCfrac{3}{+\Sp}{+\Sq}  & \shape{3}{+\Sp}{+\Sq}{+\Sk} &&  \hamps{-3}{\Sp}{\protidx{3}}{\Sq}{\qrotidx{3}}
\,\Big]
\end{alignat}
\begin{figure*}[t]
\centering
\includegraphics[scale=0.90]{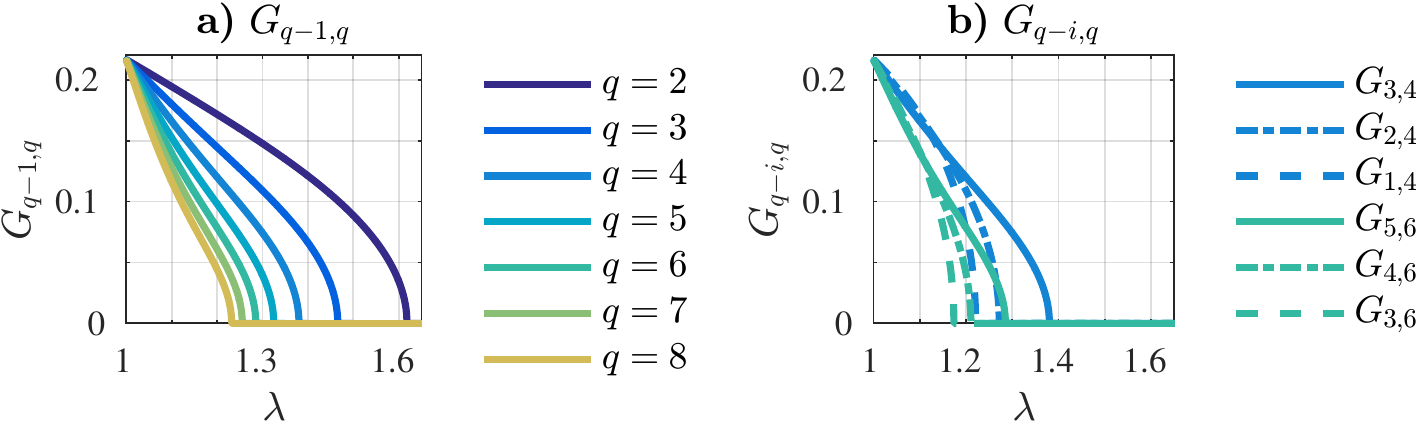}
\caption[]{Weights $\WG$. \label{fig:HNSE_shellmodel_params_appendix}}
\end{figure*}

where common terms have been factored out and the compact shell model notation $\ush{n}{\Sk}{1} = u_{\Sk}^{*}(k_n)$ adopted.
This expression is in fact a weighted sum of four helical shell models in disguise. To realise this, one needs to expand the sum over helical signs and collect terms sharing $\shapedefclean{\cdots}{}{}{\cdots}{}$ using the reflection property $\shapedefclean{-\Sp,}{-\Sq,}{-\Sk}{\cdots}{} = \shapedefclean{\Sp,}{\Sq,}{\Sk}{\cdots}{}$. Doing so and defining $\WG(\lambda) = 1/8\,Q(1,\lambda^p,\lambda^q)/(\lambda^p\lambda^q)$ the new shell model is uncovered
\begin{align}
\NSESM  \label{eqn:NSSM_compact_deriv}
\end{align}
where $\lambda,\, k_0$ are the only free parameters, $\Dissp\equiv\DisspdefSS$, and $\pqset$ are restricted by the triangle inequality $\pqveccond$.
The triad shape weight $\WG$, \submodelweight $\Wg$, and modal interaction coefficients $\Weps$ and $\Wxi$ are given by
\begin{alignat}{3}
\WG(\lambda)   &= \WGdef \label{eqn:weight_shape_deriv}\\
\Wg(\lambda)   &= \Wgdef \label{eqn:weight_interact_deriv}\\
\Weps(\lambda) &= \Wepsdef\label{eqn:weight_modalcoef_deriv}  \\
\Wxi(\lambda)  &= \Wxidef \label{eqn:weight_modalcoef_xi_deriv}. 
\end{alignat}

Lastly, a note should be made on the index notation used. The two helical sign indices $\Sp$ and $\Sq$ in $\Wg$, $\Weps$ and $\Wxi$ are separated by a comma in order to distinguish them from the single index used for modal amplitudes (such as $u_n^{\Sk\Sp}$) in which helical signs are multiplied together to produce a resulting helical sign.

\section{The interaction weights}
The functional forms of the three interaction coefficients $\Wg$, $\Weps$ and $\Wxi$ are plotted/discussed in section \ref{sec:weights} of the main text. The panels of figure \ref{fig:HNSE_shellmodel_params_appendix} show plots of the remaining weight, the triad shape weight $\WG$, ranging from the limit of local interactions $n\sim n+p \sim n+q$ to non-local $n<n+q-1 \sim n+q$ and reduced non-local $n<n+q-i \sim n+q$ for $1<i<3$. Three important results are here noticed. Firstly, the K41 assumption of local interactions being dominant is supported in panel \ref{fig:HNSE_shellmodel_params_appendix}.a. Secondly, $\WG$ is proportional to te area of the triangle formed by $\VKLEN$, $\VPLEN$ and $\VQLEN$ thereby automatically ensuring the triangle inequality is fulfilled by $\WG = 0$ if $\pqveccondinv$.
Thirdly, reducing the non-localness such that interactions tend towards coupling three different scales seems to weigh less compared to couplings involving two comparable scales, i.e. $p\sim q$ (seen from panel \ref{fig:HNSE_shellmodel_params_appendix}.b by increasing $i$).

\section{Invariants}
\label{sec:NSSM_props}

\gdef\corrsymb#1#2{\Delta_{#1,p,q}^{#2,\Sp,\Sq}}
\gdef\correlator{\corrsymb{n}{\pm} \equiv 2 k_{n-q+1}\Re[\,&\triplehelampsshort{\krotidxshifted{1}+1}{*}{+}{\protidxshifted{1}+1}{*}{\Sp}{\qrotidxshifted{1}+1}{}{\Sq} 
\mybreak 
\pm 
&
\triplehelampsshort{\krotidxshifted{1}+1}{*}{-}{\protidxshifted{1}+1}{*}{-\Sp}{\qrotidxshifted{1}+1}{}{-\Sq}]}

\gdef\Ecal{\mathpzc{E}^{\Sp,\Sq}_{p,q}(\genexpE)\,}
\gdef\Hcal{\mathpzc{H}^{\Sp,\Sq}_{p,q}(\genexpH)\,}
\gdef\Ecalshort{\mathpzc{E}^{\Sp,\Sq}_{p,q}}
\gdef\Hcalshort{\mathpzc{H}^{\Sp,\Sq}_{p,q}}

\newcommand{\tampdiff}[2]{
	\ifthenelse{\equal{#1}{1}}{
\triplehelampsshort{\protidx{1}}{*}{\Sp}{\qrotidx{1}}{}{\Sq}{\krotidx{1}}{*}{+} #2 
\triplehelampsshort{\protidx{1}}{*}{-\Sp}{\qrotidx{1}}{}{-\Sq}{\krotidx{1}}{*}{-}
	}{
	\ifthenelse{\equal{#1}{2}}{
\triplehelampsshort{\protidx{2}}{*}{\Sp}{\qrotidx{2}}{}{\Sp\Sq}{\krotidx{2}}{*}{+} #2 
\triplehelampsshort{\protidx{2}}{*}{-\Sp}{\qrotidx{2}}{}{-\Sp\Sq}{\krotidx{2}}{*}{-}
	}{
	\ifthenelse{\equal{#1}{3}}{
\triplehelampsshort{\protidx{3}}{*}{\Sq}{\qrotidx{3}}{*}{\Sp\Sq}{\krotidx{3}}{}{+} #2 
\triplehelampsshort{\protidx{3}}{*}{-\Sq}{\qrotidx{3}}{*}{-\Sp\Sq}{\krotidx{3}}{}{-}
	}{
	}}}
}

\newcommand{\tampdifftelescope}[2]{
	\ifthenelse{\equal{#1}{1}}{
\triplehelampsshort{\krotidxshifted{1}}{*}{+}{\protidxshifted{1}}{*}{\Sp}{\qrotidxshifted{1}}{}{\Sq} #2 
\triplehelampsshort{\krotidxshifted{1}}{*}{-}{\protidxshifted{1}}{*}{-\Sp}{\qrotidxshifted{1}}{}{-\Sq}
	}{
	\ifthenelse{\equal{#1}{2}}{
\triplehelampsshort{\protidxshifted{2}}{*}{\Sp}{\krotidxshifted{2}}{*}{+}{\qrotidxshifted{2}}{}{\Sp\Sq} #2 
\triplehelampsshort{\protidxshifted{2}}{*}{-\Sp}{\krotidxshifted{2}}{*}{-}{\qrotidxshifted{2}}{}{-\Sp\Sq}
	}{
	\ifthenelse{\equal{#1}{3}}{
\triplehelampsshort{\protidx{3}}{*}{\Sq}{\qrotidx{3}}{*}{\Sp\Sq}{\krotidx{3}}{}{+} #2 
\triplehelampsshort{\protidx{3}}{*}{-\Sq}{\qrotidx{3}}{*}{-\Sp\Sq}{\krotidx{3}}{}{-}
	}{
	}}}
}

\gdef\Eeqn#1{\NLderiv \Einvar{i} &=  \OuterSum \Ecalshort \sum_{\mathclap{n=1+q}}^{\mathclap{N}} 
k_{n-q}^{\genexpE}
\corrsymb{(n-1)}{-}
\label{#1}
}

\gdef\Heqn#1{\NLderiv \Hinvar{i} &=  \OuterSum \Hcalshort\sum_{\mathclap{n=1+q}}^{\mathclap{N}} 
k_{n-q}^{\genexpH}
\corrsymb{(n-1)}{+}
\label{#1}
}

\gdef\KIDXs{n}
In the helical basis energy and helicity take the simple form $E=\sum_{n=1}^{N}\modalcontr{+}$ and $H=\sum_{n=1}^{N} k_n\modalcontr{-}$ where $n=1$ and $n=N$ are the first and last shells \citep{bib:waleffe1992nature}. Here, however, we shall consider generalised quadratic invariants as in conventional shell model literature. Consider therefore the generalised energy-like and helicity-like quantities
\begin{alignat}{3}
\Einvar{i} &= \sum_{n=1}^N  k_n^{\genexpE} \modalcontr{+} 
\label{eqn:HSM_E_INVAR} \\
\Hinvar{i} &= \sum_{n=1}^N  k_n^{\genexpH} \modalcontr{-} 
\label{eqn:HSM_H_INVAR} 
\end{alignat}
where $\genexpE$ and $\genexpH$ are the $k_n$ exponents which generate $\Einvar{i}$ and $\Hinvar{i}$ respectively. 
In this notation energy is associated with index $i=1$ in $\Einvar{i}$ and has $\genexpEsymb_1 = 0$, and helicity is associated with index $i=1$ in $\Hinvar{i}$ and has $\genexpHsymb_1 = 1$.

It turns out that each of the four \submodels, here defined as the four contributions from $\sum_{\Sp,\Sq}$ in \eqnref{eqn:NSSM_compact_deriv} (main text section \ref{sec:NSSM}), inviscidly conserve energy and helicity separately for every triad shape ($\pqset$ set). Taking the time derivative of \eqnref{eqn:HSM_E_INVAR} using \eqnref{eqn:NSSM_compact_deriv} and telescoping sums by assuming a finite wave set (i.e. $u_n^\Sk = 0$ for $n<1$ and $n>N$), one finds the non-linear (N.L.) rate-of-change of $\Einvar{i}$ is given by the long but straight forward calculation
\begin{alignat}{5}
\NLderiv \Einvar{i} &=  
\sum_{n=1}^N k_n^{\genexpE} (
\ush{n}{+}{1}d_t\ush{n}{+}{0} 
+ \ush{n}{-}{1}d_t\ush{n}{-}{0} 
) 
+ c.c. 
\notag\\
&= \OuterSum\sum_{\mathclap{n=1+q}}^{\mathclap{N}}
k_{n-q}^{\genexpE+1}
\Big[
\mybreak
\Big(& \tampdifftelescope{1}{\;\;-} \Big) \notag\\
-(\genE{\genexpE})^p\Weps \Big(&\tampdifftelescope{2}{-} \Big) \notag\\
+(\genE{\genexpE})^q\Wxi \Big(& \tampdifftelescope{3}{\;\;\,-} \Big) \Big] \mybreak & \hspace{10em} + c.c. \label{eqn:calc_invariants}
\end{alignat}
From here it is noticed that the second and third velocity triple-product differences are equal to the first times $\Sp$ and $\Sq$ respectively, thus allowing to be factored out.
A similar calculation may be done for $\Hinvar{i}$ yielding a positive sign between the velocity triple-products,
implying all three triple-product differences are similar.
Tidying up by defining correlators as
\begin{align}
\correlator
\end{align}
the energy and helicity equations become 
\begin{alignat}{5}
\Eeqn{eqn:NSSM_E_eqn_deriv}\\
\Heqn{eqn:NSSM_H_eqn_deriv} 
\end{alignat}
where 
\begin{alignat}{5}
\Ecal &=\Egeneqn{p}{q}{\genexpE} \label{eqn:Ecal}\\
\Hcal &=\Hgeneqn{p}{q}{\genexpH}. \label{eqn:Hcal}
\end{alignat}
From these equations it is apparent that conservation of $\Einvar{i}$ and $\Hinvar{i}$ requires $\Ecalshort=0$ and $\Hcalshort=0$.
Plugging $\genexpE = \genexpEsymb_1 = 0$ into $\Ecalshort$ one finds energy is always conserved independently of triad shape ($\pqset$ pair) and \submodel ($\SpSqset$ pair).
Other solutions to $\Ecalshort=0$, however, depend on the specific triad shapes by $\pqset$. Since these roots are unlikely to be shared across triad shapes, the remaining invariants can be considered triad shape-specific invariants or \emph{pseudo-invariants} because they are broken when mixing triad shapes. 
Furthermore, the solutions related to a given triad shape vary between the four \submodels, implying mixing \submodels also break the energy pseudo-invariants. In this light energy pseudo-invariants can be considered artefacts from not resolving the system properly by reducing the set of triadic interactions to only one triad shape and one \submodel.

In a similar fashion each \submodel inviscidly conserves helicity ($\genexpH = \genexpHsymb_1 = 1$) separately for every triad shape since $\Hcalshort = 0$
by substituting \eqnref{eqn:weight_modalcoef_deriv} and \eqnref{eqn:weight_modalcoef_xi_deriv} in.
The remaining helicity-like invariants behave similarly to the energy-like invariants and are thus denoted helicity pseudo-invariants.

\section{Spectral fluxes}
\label{sec:NSSM_fluxes}

\gdef\KIDXs{m}

\gdef\Efluxeqn{\Eflux{i}{n} &= \OuterSum \Big[ 
\sum_{\mathclap{m=n+1}}^{\mathclap{n+q}} k_{m-q}^{\genexpE} \corrsymb{(\KIDXs-1)}{-} 
\mybreak & \myqquad\myqquad
-\Sp\Weps\sum_{\mathclap{m=n+1}}^{\mathclap{n+q-p}} k_{m-q+p}^{\genexpE} \corrsymb{(\KIDXs-1)}{-} 
\Big]}

\gdef\Hfluxeqn{\Hflux{i}{n} &= \OuterSum \Big[ 
\sum_{\mathclap{m=n+1}}^{\mathclap{n+q}} k_{m-q}^{\genexpH} \corrsymb{(\KIDXs-1)}{+} 
\mybreak & \myqquad\myqquad\phantom{\Sp}
-\Weps\sum_{\mathclap{m=n+1}}^{\mathclap{n+q-p}} k_{m-q+p}^{\genexpH} \corrsymb{(\KIDXs-1)}{+} 
\Big]}

Non-linear spectral fluxes of $\Einvar{i}$ and $\Hinvar{i}$ through the $n$-th shell are given as the transfers from all wave numbers less than $k_n$ to wave numbers larger than, that is $\Eflux{i}{n} = \NLderiv\sum_{m=1}^n k_m^{\genexpE} \modalcontr{+} $ and $\Hflux{i}{n} = \NLderiv\sum_{m=1}^n k_m^{\genexpH} \modalcontr{-} $. 
Following the calculations through one finds \eqnref{eqn:calc_invariants} becomes (breaking the sum at $n$ instead of $N$)

\gdef\KIDX{m} 
\begin{alignat}{5}
\Eflux{i}{n} 
=
\OuterSum \Big[ 
\Ecalshort\sum_{\mathclap{m=1+q}}^{\mathclap{n}} 
k_{m-q}^{\genexpE}
\corrsymb{(\KIDX-1)}{-}
\mybreak 
+\sum_{\mathclap{m=n+1}}^{\mathclap{n+q}} 
k_{m-q}^{\genexpE} 
\corrsymb{(\KIDX-1)}{-}
-\Sp
\Weps\sum_{\mathclap{m=n+1}}^{\mathclap{n+q-p}} 
k_{m-q+p}^{\genexpE} 
\corrsymb{(\KIDX-1)}{-}
\Big] 
\label{eqn:NSSM_E_flux_deriv_1}
\end{alignat}
where summation over the shared range $1+q \leq m \leq n$ has been grouped together into the first term.
This term, however, must clearly vanishes since $\Ecal=0$ is required for $\Einvar{i}$ to be an invariant. 
Going through similar calculations for $\Hflux{i}{n}$, one finally finds 
\gdef\KIDX{n}
\gdef\KIDX{n} 
\begin{alignat}{5}
\Efluxeqn \label{eqn:NSSM_E_flux_deriv}\\
\Hfluxeqn \label{eqn:NSSM_H_flux_deriv}
\end{alignat}
which, taking into account the different correlator definitions, conforms with equivalent expressions for helically decomposed GOY and Sabra models.

	\bibliographystyle{apsrev4-1}\bibliography{helical_shell_models}
\else
	
	\twocolumngrid	
	\bibliographystyle{apsrev4-1}\bibliography{helical_shell_models}

\begin{thebibliography}{16}%
\makeatletter
\providecommand \@ifxundefined [1]{%
 \@ifx{#1\undefined}
}%
\providecommand \@ifnum [1]{%
 \ifnum #1\expandafter \@firstoftwo
 \else \expandafter \@secondoftwo
 \fi
}%
\providecommand \@ifx [1]{%
 \ifx #1\expandafter \@firstoftwo
 \else \expandafter \@secondoftwo
 \fi
}%
\providecommand \natexlab [1]{#1}%
\providecommand \enquote  [1]{``#1''}%
\providecommand \bibnamefont  [1]{#1}%
\providecommand \bibfnamefont [1]{#1}%
\providecommand \citenamefont [1]{#1}%
\providecommand \href@noop [0]{\@secondoftwo}%
\providecommand \href [0]{\begingroup \@sanitize@url \@href}%
\providecommand \@href[1]{\@@startlink{#1}\@@href}%
\providecommand \@@href[1]{\endgroup#1\@@endlink}%
\providecommand \@sanitize@url [0]{\catcode `\\12\catcode `\$12\catcode
  `\&12\catcode `\#12\catcode `\^12\catcode `\_12\catcode `\%12\relax}%
\providecommand \@@startlink[1]{}%
\providecommand \@@endlink[0]{}%
\providecommand \url  [0]{\begingroup\@sanitize@url \@url }%
\providecommand \@url [1]{\endgroup\@href {#1}{\urlprefix }}%
\providecommand \urlprefix  [0]{URL }%
\providecommand \Eprint [0]{\href }%
\providecommand \doibase [0]{http://dx.doi.org/}%
\providecommand \selectlanguage [0]{\@gobble}%
\providecommand \bibinfo  [0]{\@secondoftwo}%
\providecommand \bibfield  [0]{\@secondoftwo}%
\providecommand \translation [1]{[#1]}%
\providecommand \BibitemOpen [0]{}%
\providecommand \bibitemStop [0]{}%
\providecommand \bibitemNoStop [0]{.\EOS\space}%
\providecommand \EOS [0]{\spacefactor3000\relax}%
\providecommand \BibitemShut  [1]{\csname bibitem#1\endcsname}%
\let\auto@bib@innerbib\@empty
\bibitem [{\citenamefont {Waleffe}(1992)}]{bib:waleffe1992nature}%
  \BibitemOpen
  \bibfield  {author} {\bibinfo {author} {\bibfnamefont {F.}~\bibnamefont
  {Waleffe}},\ }\href@noop {} {\bibfield  {journal} {\bibinfo  {journal}
  {Physics of Fluids A: Fluid Dynamics (1989-1993)}\ }\textbf {\bibinfo
  {volume} {4}},\ \bibinfo {pages} {350} (\bibinfo {year} {1992})}\BibitemShut
  {NoStop}%
\bibitem [{\citenamefont {Biferale}\ \emph {et~al.}(2012)\citenamefont
  {Biferale}, \citenamefont {Musacchio},\ and\ \citenamefont
  {Toschi}}]{bib:biferale2012inverse}%
  \BibitemOpen
  \bibfield  {author} {\bibinfo {author} {\bibfnamefont {L.}~\bibnamefont
  {Biferale}}, \bibinfo {author} {\bibfnamefont {S.}~\bibnamefont {Musacchio}},
  \ and\ \bibinfo {author} {\bibfnamefont {F.}~\bibnamefont {Toschi}},\
  }\href@noop {} {\bibfield  {journal} {\bibinfo  {journal} {Physical review
  letters}\ }\textbf {\bibinfo {volume} {108}},\ \bibinfo {pages} {164501}
  (\bibinfo {year} {2012})}\BibitemShut {NoStop}%
\bibitem [{\citenamefont {Gledzer}(1973)}]{bib:gledzer1973system}%
  \BibitemOpen
  \bibfield  {author} {\bibinfo {author} {\bibfnamefont {E.}~\bibnamefont
  {Gledzer}},\ }in\ \href@noop {} {\emph {\bibinfo {booktitle} {Soviet Physics
  Doklady}}},\ Vol.~\bibinfo {volume} {18}\ (\bibinfo {year} {1973})\ p.\
  \bibinfo {pages} {216}\BibitemShut {NoStop}%
\bibitem [{\citenamefont {L’vov}\ \emph {et~al.}(1998)\citenamefont
  {L’vov}, \citenamefont {Podivilov}, \citenamefont {Pomyalov}, \citenamefont
  {Procaccia},\ and\ \citenamefont {Vandembroucq}}]{bib:sabra_lvov}%
  \BibitemOpen
  \bibfield  {author} {\bibinfo {author} {\bibfnamefont {V.~S.}\ \bibnamefont
  {L’vov}}, \bibinfo {author} {\bibfnamefont {E.}~\bibnamefont {Podivilov}},
  \bibinfo {author} {\bibfnamefont {A.}~\bibnamefont {Pomyalov}}, \bibinfo
  {author} {\bibfnamefont {I.}~\bibnamefont {Procaccia}}, \ and\ \bibinfo
  {author} {\bibfnamefont {D.}~\bibnamefont {Vandembroucq}},\ }\href@noop {}
  {\bibfield  {journal} {\bibinfo  {journal} {Physical Review E}\ }\textbf
  {\bibinfo {volume} {58}},\ \bibinfo {pages} {1811} (\bibinfo {year}
  {1998})}\BibitemShut {NoStop}%
\bibitem [{\citenamefont {Ditlevsen}(2000)}]{bib:sabra_ditlevsen}%
  \BibitemOpen
  \bibfield  {author} {\bibinfo {author} {\bibfnamefont {P.~D.}\ \bibnamefont
  {Ditlevsen}},\ }\href@noop {} {\bibfield  {journal} {\bibinfo  {journal}
  {Physical Review E}\ }\textbf {\bibinfo {volume} {62}},\ \bibinfo {pages}
  {484} (\bibinfo {year} {2000})}\BibitemShut {NoStop}%
\bibitem [{\citenamefont {Biferale}\ and\ \citenamefont
  {Kerr}(1995)}]{bib:biferale1995role}%
  \BibitemOpen
  \bibfield  {author} {\bibinfo {author} {\bibfnamefont {L.}~\bibnamefont
  {Biferale}}\ and\ \bibinfo {author} {\bibfnamefont {R.~M.}\ \bibnamefont
  {Kerr}},\ }\href@noop {} {\bibfield  {journal} {\bibinfo  {journal} {Physical
  Review E}\ }\textbf {\bibinfo {volume} {52}},\ \bibinfo {pages} {6113}
  (\bibinfo {year} {1995})}\BibitemShut {NoStop}%
\bibitem [{\citenamefont {Benzi}\ \emph {et~al.}(1996)\citenamefont {Benzi},
  \citenamefont {Biferale}, \citenamefont {Kerr},\ and\ \citenamefont
  {Trovatore}}]{bib:benzi1996helical}%
  \BibitemOpen
  \bibfield  {author} {\bibinfo {author} {\bibfnamefont {R.}~\bibnamefont
  {Benzi}}, \bibinfo {author} {\bibfnamefont {L.}~\bibnamefont {Biferale}},
  \bibinfo {author} {\bibfnamefont {R.~M.}\ \bibnamefont {Kerr}}, \ and\
  \bibinfo {author} {\bibfnamefont {E.}~\bibnamefont {Trovatore}},\ }\href@noop
  {} {\bibfield  {journal} {\bibinfo  {journal} {Physical Review E}\ }\textbf
  {\bibinfo {volume} {53}},\ \bibinfo {pages} {3541} (\bibinfo {year}
  {1996})}\BibitemShut {NoStop}%
\bibitem [{\citenamefont {Biferale}\ \emph
  {et~al.}(1998{\natexlab{a}})\citenamefont {Biferale}, \citenamefont
  {Pierotti},\ and\ \citenamefont {Toschi}}]{bib:biferale1998helicityA}%
  \BibitemOpen
  \bibfield  {author} {\bibinfo {author} {\bibfnamefont {L.}~\bibnamefont
  {Biferale}}, \bibinfo {author} {\bibfnamefont {D.}~\bibnamefont {Pierotti}},
  \ and\ \bibinfo {author} {\bibfnamefont {F.}~\bibnamefont {Toschi}},\
  }\href@noop {} {\bibfield  {journal} {\bibinfo  {journal} {Physical Review
  E}\ }\textbf {\bibinfo {volume} {57}},\ \bibinfo {pages} {R2515} (\bibinfo
  {year} {1998}{\natexlab{a}})}\BibitemShut {NoStop}%
\bibitem [{\citenamefont {Biferale}\ \emph
  {et~al.}(1998{\natexlab{b}})\citenamefont {Biferale}, \citenamefont
  {Pierotti},\ and\ \citenamefont {Toschi}}]{bib:biferale1998helicityB}%
  \BibitemOpen
  \bibfield  {author} {\bibinfo {author} {\bibfnamefont {L.}~\bibnamefont
  {Biferale}}, \bibinfo {author} {\bibfnamefont {D.}~\bibnamefont {Pierotti}},
  \ and\ \bibinfo {author} {\bibfnamefont {F.}~\bibnamefont {Toschi}},\
  }\href@noop {} {\bibfield  {journal} {\bibinfo  {journal} {Le Journal de
  Physique IV}\ }\textbf {\bibinfo {volume} {8}},\ \bibinfo {pages} {Pr6}
  (\bibinfo {year} {1998}{\natexlab{b}})}\BibitemShut {NoStop}%
\bibitem [{\citenamefont {{De Pietro}}\ \emph {et~al.}(2015)\citenamefont {{De
  Pietro}}, \citenamefont {{Biferale}},\ and\ \citenamefont
  {{Mailybaev}}}]{bib:DePietro2015arXiv150806390D}%
  \BibitemOpen
  \bibfield  {author} {\bibinfo {author} {\bibfnamefont {M.}~\bibnamefont {{De
  Pietro}}}, \bibinfo {author} {\bibfnamefont {L.}~\bibnamefont {{Biferale}}},
  \ and\ \bibinfo {author} {\bibfnamefont {A.~A.}\ \bibnamefont
  {{Mailybaev}}},\ }\href@noop {} {\bibfield  {journal} {\bibinfo  {journal}
  {ArXiv e-prints}\ } (\bibinfo {year} {2015})},\ \Eprint
  {http://arxiv.org/abs/1508.06390} {arXiv:1508.06390 [physics.flu-dyn]}
  \BibitemShut {NoStop}%
\bibitem [{\citenamefont {Biferale}\ \emph {et~al.}(2013)\citenamefont
  {Biferale}, \citenamefont {Musacchio},\ and\ \citenamefont
  {Toschi}}]{bib:biferale2013split}%
  \BibitemOpen
  \bibfield  {author} {\bibinfo {author} {\bibfnamefont {L.}~\bibnamefont
  {Biferale}}, \bibinfo {author} {\bibfnamefont {S.}~\bibnamefont {Musacchio}},
  \ and\ \bibinfo {author} {\bibfnamefont {F.}~\bibnamefont {Toschi}},\
  }\href@noop {} {\bibfield  {journal} {\bibinfo  {journal} {Journal of Fluid
  Mechanics}\ }\textbf {\bibinfo {volume} {730}},\ \bibinfo {pages} {309}
  (\bibinfo {year} {2013})}\BibitemShut {NoStop}%
\bibitem [{\citenamefont {{Sahoo}}\ \emph {et~al.}(2015)\citenamefont
  {{Sahoo}}, \citenamefont {{Bonaccorso}},\ and\ \citenamefont
  {{Biferale}}}]{bib:sahoo2015}%
  \BibitemOpen
  \bibfield  {author} {\bibinfo {author} {\bibfnamefont {G.}~\bibnamefont
  {{Sahoo}}}, \bibinfo {author} {\bibfnamefont {F.}~\bibnamefont
  {{Bonaccorso}}}, \ and\ \bibinfo {author} {\bibfnamefont {L.}~\bibnamefont
  {{Biferale}}},\ }\href@noop {} {\bibfield  {journal} {\bibinfo  {journal}
  {ArXiv e-prints}\ } (\bibinfo {year} {2015})},\ \Eprint
  {http://arxiv.org/abs/1506.04906} {arXiv:1506.04906 [nlin.CD]} \BibitemShut
  {NoStop}%
\bibitem [{\citenamefont {Ditlevsen}\ and\ \citenamefont
  {Mogensen}(1996)}]{bib:ditlevsenmogensen1996}%
  \BibitemOpen
  \bibfield  {author} {\bibinfo {author} {\bibfnamefont {P.~D.}\ \bibnamefont
  {Ditlevsen}}\ and\ \bibinfo {author} {\bibfnamefont {I.~A.}\ \bibnamefont
  {Mogensen}},\ }\href@noop {} {\bibfield  {journal} {\bibinfo  {journal}
  {Phys. Rev. E}\ }\textbf {\bibinfo {volume} {53}},\ \bibinfo {pages} {4785}
  (\bibinfo {year} {1996})}\BibitemShut {NoStop}%
\bibitem [{\citenamefont {Kraichnan}\ and\ \citenamefont
  {Montgomery}(1980)}]{bib:robert1980two}%
  \BibitemOpen
  \bibfield  {author} {\bibinfo {author} {\bibfnamefont {R.}~\bibnamefont
  {Kraichnan}}\ and\ \bibinfo {author} {\bibfnamefont {D.}~\bibnamefont
  {Montgomery}},\ }\href@noop {} {\bibfield  {journal} {\bibinfo  {journal}
  {Rep. Prog. Phys}\ }\textbf {\bibinfo {volume} {43}} (\bibinfo {year}
  {1980})}\BibitemShut {NoStop}%
\bibitem [{\citenamefont {Brissaud}\ \emph {et~al.}(1973)\citenamefont
  {Brissaud}, \citenamefont {Frisch}, \citenamefont {Leorat}, \citenamefont
  {Lesieur},\ and\ \citenamefont {Mazure}}]{bib:brissaud1973helicity}%
  \BibitemOpen
  \bibfield  {author} {\bibinfo {author} {\bibfnamefont {A.}~\bibnamefont
  {Brissaud}}, \bibinfo {author} {\bibfnamefont {U.}~\bibnamefont {Frisch}},
  \bibinfo {author} {\bibfnamefont {J.}~\bibnamefont {Leorat}}, \bibinfo
  {author} {\bibfnamefont {M.}~\bibnamefont {Lesieur}}, \ and\ \bibinfo
  {author} {\bibfnamefont {A.}~\bibnamefont {Mazure}},\ }\href@noop {}
  {\bibfield  {journal} {\bibinfo  {journal} {Physics of Fluids (1958-1988)}\
  }\textbf {\bibinfo {volume} {16}},\ \bibinfo {pages} {1366} (\bibinfo {year}
  {1973})}\BibitemShut {NoStop}%
\bibitem [{\citenamefont {Gilbert}\ \emph {et~al.}(2002)\citenamefont
  {Gilbert}, \citenamefont {L’vov}, \citenamefont {Pomyalov},\ and\
  \citenamefont {Procaccia}}]{bib:gilbert2002inverse}%
  \BibitemOpen
  \bibfield  {author} {\bibinfo {author} {\bibfnamefont {T.}~\bibnamefont
  {Gilbert}}, \bibinfo {author} {\bibfnamefont {V.~S.}\ \bibnamefont
  {L’vov}}, \bibinfo {author} {\bibfnamefont {A.}~\bibnamefont {Pomyalov}}, \
  and\ \bibinfo {author} {\bibfnamefont {I.}~\bibnamefont {Procaccia}},\
  }\href@noop {} {\bibfield  {journal} {\bibinfo  {journal} {Physical review
  letters}\ }\textbf {\bibinfo {volume} {89}},\ \bibinfo {pages} {074501}
  (\bibinfo {year} {2002})}\BibitemShut {NoStop}%
\end{thebibliography}%
\fi

\end{document}